\documentclass{aa}

\usepackage{graphicx}
\usepackage{txfonts}
\usepackage{xcolor}

\usepackage[]{hyperref}

\makeatletter
\renewcommand*\aa@pageof{, page \thepage{} of \pageref*{LastPage}}
\makeatother

\defcitealias{Carretta2009u}{C09u}
\defcitealias{Carretta2009g}{C09g}
\defcitealias{Schiappacasse-Ulloa2024}{JSU24}
\newcommand{\teff}{T$_{\mathrm{eff}}$~}

\usepackage{adjustbox}
\usepackage{lscape}
\usepackage{longtable}
\usepackage{subcaption}

\begin{document}

   \title{Intrinsic iron abundance spreads in globular clusters}

   \subtitle{}

   \author{J. Schiappacasse-Ulloa
          \inst{1}
          \and
          S. Lucatello\inst{2}
          \and
          L. Magrini\inst{1}
          \and
          A. Bragaglia\inst{3}
          \and
          E. Carretta\inst{3}}

   \institute{INAF–Osservatorio Astrofisico di Arcetri,  Largo Enrico Fermi 5, 50125 Florence, Italy\\
              \email{jose.schiappacasse@inaf.it}
         \and
             INAF–Osservatorio Astronomico di Padova, Vicolo dell’Osservatorio 5, 35122 Padova, Italy
        \and
             INAF-Osservatorio di Astrofisica e Scienza dello Spazio, via P. Gobetti 93/2, 40129 Bologna, Italy
             }

   \date{Received September 15, 1996; accepted March 16, 1997}

  \abstract
   {Globular clusters host multiple stellar populations, likely composed of a subset of first-generation stars that enriched the intracluster medium and gave rise to second-generation (SG) stars, each characterised by distinctive chemical patterns. These patterns typically include enhancements in elements such as N, Na, and Al, coupled with depletions in C, O, and Mg in SG stars. Traditionally, heavier elements such as those in the iron peak were considered unaffected in most clusters. However, recent studies have reported significant internal spreads in these elements, suggesting a more complex picture of chemical enrichment within globular clusters.} 
   {This study seeks to derive precise and homogeneous differential iron abundances in a large sample of globular clusters. By doing so, our aim is to investigate the presence of intrinsic iron spreads within them and to assess the existence of differences in Fe between their stellar populations.}
   {We used the Python-based tool \texttt{Q2} to determine both differential stellar parameters and iron abundances for 92 sibling stars -- defined by the similarities in their stellar parameters -- across 13 Galactic globular clusters. This differential approach reduces the influence of non-local thermodynamic equilibrium effects, and minimises observational errors, and systematic biases linked to stellar parameters. We performed Monte Carlo simulations to evaluate the statistical significance of the measured spreads.}
   {Most of the globular clusters in our sample do not show evidence of statistically significant iron spreads. Only a few exceptions emerge, namely NGC~1851, NGC~3201, and NGC~5634, which display a highly significant iron spread. In particular, NGC~3201 shows a particularly pronounced spread in its first-generation population, reflecting a potential inhomogeneous iron abundance in its pristine material. Finally, through statistical tests, we conclude that our data do not support the presence of a widespread iron variation in globular clusters.}
   {}


   \maketitle

\section{Introduction}
\label{introduction}

Globular clusters (GCs) serve as key tracers of the formation and evolution of the Galactic halo \citep{Searle1978,Zinn1985,Carretta2010global,Martell2011}, and they also trace the formation history of the Galactic bulge \citep{Lee2019}. Over the past few decades, these ancient stellar systems have attracted sustained attention following the discovery of the multiple stellar population (MP) phenomenon, which has fundamentally reshaped our understanding of their nature and formation history \citep{Gratton2004,Gratton2012,Bastian2018}.

Several scenarios have been proposed to explain the origin of MPs. In particular, one of the proposed models suggests that a subset of first-generation (FG) stars evolved and polluted the intracluster medium with material processed through hot H-burning \citep{Gratton2019,Bastian2018}. This chemically enriched material, mixed with pristine gas, subsequently formed second-generation (SG) stars. Other proposed models to explain this phenomenon do not even require multiple episodes of star formation. For example, the inertial-inflow mechanism scaled up to GCs \citep{Gieles2025} describes a single burst of star formation driven by an accreting extremely massive star (10$^{3}$-10$^{4}$M$_\sun$). Over a brief 1-2 Myr window, these stars accrete pristine gas and eject chemically processed material via stellar winds. Consequently, low-mass stars that continuously formed throughout this period incorporate varying fractions of polluted material, naturally producing the observed spectrum of stellar populations.\footnote{It is worth noting that throughout this paper we use the labels FG and SG solely as a convenient nomenclature and do not intend them to imply any specific formation scenario for MPs.} 
As a consequence, present-day Galactic GCs host at least two chemically distinct stellar populations. The FG stars exhibit abundance patterns similar to those observed in field stars at the same metallicity, whereas the SG stars display enhanced abundances of elements such as N, Na, and Al, together with depleted abundances of C, O, and Mg. Spectroscopic studies have extensively characterised these chemical peculiarities in MPs by providing direct measurements of elemental abundances, particularly for elements involved in hot H-burning, such as O, Na, Mg, and Al \citep[e.g.][]{Johnson2005,Carretta2009u,Carretta2009g,Meszaros2015,Pancino2017}, but also Li \citep{SchiappacasseUlloa2021}, and neutron-capture elements \citep{Schiappacasse-Ulloa2023}.

In contrast to the significant variations observed in light elements, with the exception of a select number of clusters (e.g. M15, M22), data show no evidence for significant spreads in heavy elements \citep[see e.g.][]{Schiappacasse-Ulloa2025}. Variations in iron-peak elements (e.g. Fe and Ni) beyond observational uncertainties appear only in a limited number of GCs (e.g. $\omega$-Cen \cite{Johnson2010}, M~54 \cite{Carretta2010}, and NGC~2808 \cite{Lardo2023}), while most studied clusters show no evidence of significant iron spreads \citep[see e.g.][]{Bailin2019}. This chemical uniformity suggests that Type Ia supernovae (SNe Ia), which represent major producers of iron, did not significantly contribute to the chemical enrichment of SG stars. Two main arguments support this interpretation: (i) SNe Ia typically explode on timescales longer than the formation timescale of SG stars, and (ii) their highly energetic ejecta would likely exceed the escape velocity of most GCs, preventing efficient retention of iron-rich material. Consequently, any proposed polluter must alter the light-element chemistry without modifying the overall iron content of the cluster.

Photometric studies that exploit flux differences across specific spectral regions, however, argue for a more complex picture. In this context, \citet{Milone2017} formalised the concept of the chromosome map (ChM), a pseudo–colour–colour diagram constructed from combinations of Hubble Space Telescope (HST) filters designed to identify and characterise MPs. Based on the stellar distributions observed in the ChM, \citet{Milone2017} classified GCs into two categories. Type I clusters ($\sim$80$\%$ of Galactic GCs) display two well-defined stellar sequences and generally lack intrinsic iron abundance variations. In contrast, Type II clusters exhibit more complex ChMs, characterised by an additional stellar component located redwards of the typical SG population. In some systems, this feature has been linked to internal iron variations; however, spectroscopic analyses of several photometrically defined Type II clusters -- for example, NGC~6388 \citet{Carretta2022} and NGC~362 \citet{Carretta2013,Meszaros2020} -- indicate that any iron variation remains negligible.

However, Type II clusters display additional complexity in both their photometric and spectroscopic properties. For example, NGC~5139 ($\omega$-Cen) hosts multiple distinct stellar populations spanning a wide metallicity range, while NGC~6715 (M~54) exhibits a substantial [Fe/H] spread and likely formed within the Sagittarius Dwarf Spheroidal Galaxy, as recent dynamical studies confirm \citep{AlfaroCuello2020,Kacharov2022}\footnote{It is worth noticing that both $\omega$-Cen and M~54 are widely interpreted as nuclear star clusters of accreted dwarf galaxies rather than genuine mono-metallic Galactic GCs.}. Their properties therefore suggest that at least some systems that exhibit internal iron abundance variations may have an extragalactic origin.

Adding further complexity, \citet{Legnardi2022} analysed HST photometry for 55 GCs and reported significant internal iron spreads, in several cases exceeding 0.1 dex and reaching values above 0.2 dex, in a surprisingly large fraction of the sample, including several Type I clusters. Moreover, these intriguing iron variations were found among FG stars. This result challenges the prevailing paradigm and points towards a more complex enrichment history.

Some spectroscopic studies support the findings of \citet{Legnardi2022} in a limited number of clusters. For example, \citet{Lardo2022} used high-resolution spectroscopy to report significant iron variations, while \citet{Latour2025} analysed MUSE spectra and also detected internal Fe spreads. However, the iron variations reported by \citet{Latour2025} are generally smaller than those inferred by \citet{Legnardi2022} for FG stars. In addition, \citet{Latour2025} detected iron spreads among SG stars, although these remain less pronounced than those observed in the FG population. In contrast, other spectroscopic studies find no evidence for significant iron variations \citep[e.g.][]{Carretta2025}. In particular, \citet{Carretta2025inc} highlighted a discrepancy between the iron spreads inferred from the photometric analysis of \citet{Legnardi2022} and those derived from spectroscopic observations. Furthermore, \citet{Carretta2025} demonstrated that the uncertainties associated with iron abundances derived from traditional high-resolution spectroscopic analyses of individual stars (i.e. through spectral synthesis or equivalent-width measurements) are comparable to the magnitude of the iron spreads inferred from photometric studies, such as those reported by \citet{Marino2019} for NGC~3201 and by \citet{Marino2023} for NGC~104.

To overcome the limitations inherent in traditional spectroscopic methods, we explore the use of differential abundance analysis, which enables significantly higher precision in the determination of elemental abundances \citep{Ramirez2009}. This technique is applicable only to pairs or groups of stars with nearly identical atmospheric parameters and its use has, so far, been limited to a small number of cases in the context of GCs \citep[see e.g.][]{Monty2023,Lardo2023}. Nevertheless, differential analysis provides the precision required to detect subtle internal chemical variations in GCs.

In this work, we perform a differential abundance analysis of 92 stars with similar stellar parameters across 13 GCs, with the aim of quantifying variations in their iron content. This dataset represents the largest sample of stars in GC analysed with this technique to date and enables a robust assessment of subtle iron abundance variations within clusters. 

The paper is organised as follows. Section~\ref{Sec: Samp-Meth} describes the sample selection and methodology. Section~\ref{Sec: CompLite} presents a comparison with the results available in the literature. Section~\ref{Sec: Results} presents our findings, Section~\ref{Sec: Discussion} discusses their implications, and Section~\ref{Sec: Conclusions} summarises the main conclusions and outlines the broader significance of this work.

\section{Methodology}
\label{Sec: Samp-Meth}

\subsection{Sample}

We based our analysis on the stellar sample of \citet[][hereafter JSU24]{Schiappacasse-Ulloa2024}, based on high-resolution spectra  obtained with the Ultraviolet and Visual Echelle Spectrograph \citep[UVES,][]{Dekker2000SPIE.4008..534D} mounted on the Very Large Telescope (VLT) of the European Southern Observatory (ESO) with a resolving power of R = 47,000, a spectral range of 480.0–680.0 nm, and a typical signal-to-noise of $\sim$85. These data were originally presented by \citet{Carretta2009u} and later extended to the GCs NGC~1851 and NGC~5634 by \citet{Carretta2011} and \citet{Carretta2017}, respectively. While \citet{Carretta2009u} focused primarily on light-element abundances, \citet{Carretta2011} and \citet{Carretta2017} expanded the analysis to include heavy elements for a subset of these clusters. Together, these studies provide homogeneous stellar parameters and elemental abundances for more than 200 red giant branch (RGB) stars across 18 GCs, spanning a wide range of metallicities. 
\citet{Carretta2009u} derived photometric \teff in a homogeneous manner, and \citet{Carretta2015} applied the same methodology to NGC~2808, ensuring consistency with the earlier analysis. In particular, the initial estimates of \teff were obtained from the V-K colour–temperature calibration of \citet{Alonso1999}. The authors then refined these values using an empirical relation between \teff and V magnitude (or K magnitude in clusters affected by significant reddening), derived from a subsample of well-behaved stars. We refer the reader to \citet{Carretta2009u} for more details.

\begin{figure*}
        \centering
        \includegraphics[width=0.9\textwidth]{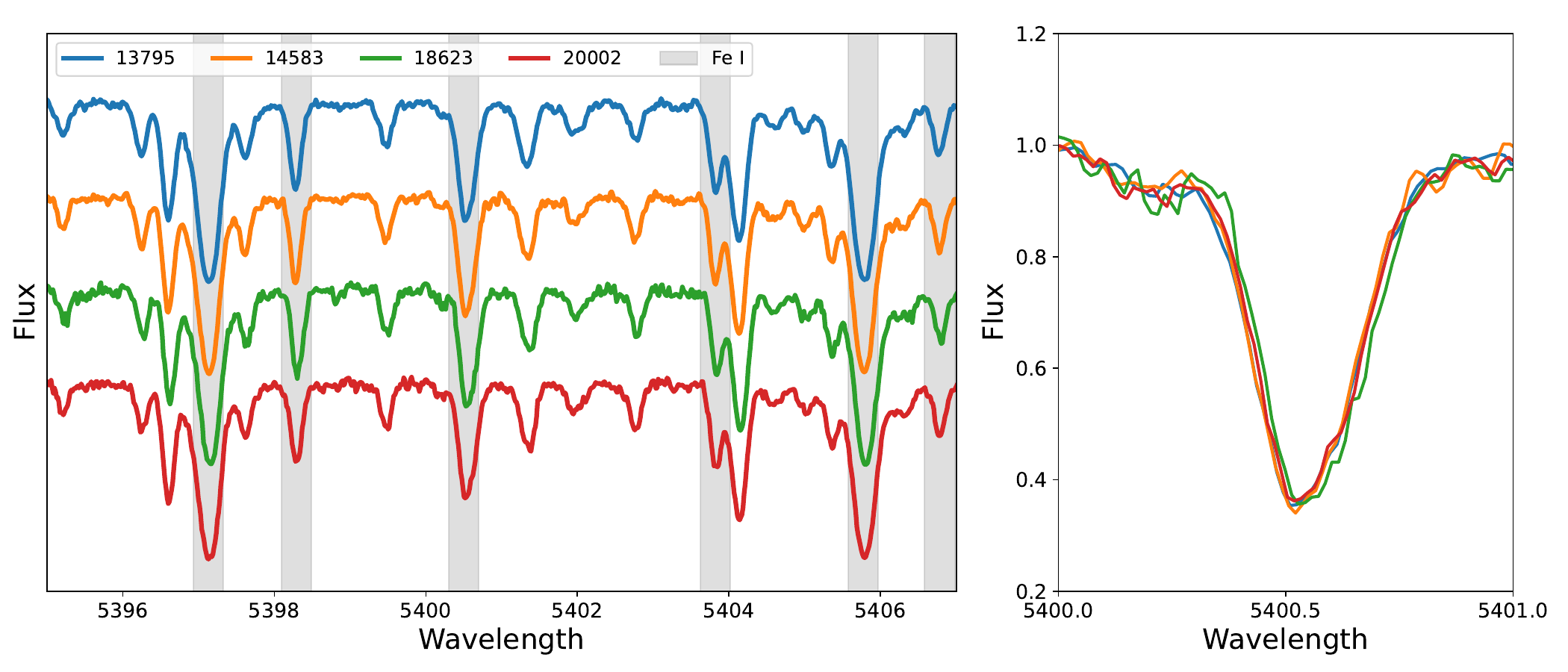}
        \caption{Spectral region containing Fe I lines (shaded areas). The different coloured curves correspond to stars in the GC NGC 104 with nearly identical stellar parameters. Star 13795 (blue curve) belongs to the FG, whereas the remaining stars belong to the SG. The right-hand panel shows a zoomed view of one of the Fe I lines.}
        \label{fig:comp_lines}
\end{figure*}

Although the original abundance analyses were performed with rigorous methodologies, the quoted uncertainties are comparable to the subtle abundance variations that we aim to detect. This limitation becomes evident when examining stars with nearly identical atmospheric parameters. For illustrative purposes, Fig.~\ref{fig:comp_lines} shows, in the left panel, a set of Fe I lines for a group of such stars in NGC~104. The right panel zooms in on a representative line, highlighting subtle differences in line depth between stars. These differences may reflect genuine variations in iron abundance that are, however, partially masked by the uncertainties reported in the original studies.

To overcome this limitation, we reanalysed the same spectra using a strictly differential abundance approach. This method minimises systematic effects and significantly reduces uncertainties (see below), thereby increasing sensitivity to intrinsic abundance spreads within GCs. We describe the methodology in detail in the following section.

\subsection{Method}

To investigate potential intra-cluster abundance variations within each GC, we carried out a differential chemical analysis on groups of stars with similar stellar parameters -- hereafter referred to as sibling stars -- using a line-by-line abundance comparison. By selecting stars with nearly identical atmospheric parameters (within $\pm$100 K), we minimise systematic trends associated with \teff, surface gravity ($\log$ g), and microturbulence ($\nu_t$). Although the Fe lines in our sample may be affected by departures from local thermodynamic equilibrium (LTE; \citealt{Lind2012}), these effects remain negligible when comparing stars with such small parameter differences and small abundance differences, as they influence all members of a group in nearly the same way \citep{Yong2013}. The differential method also suppresses several other major sources of systematic uncertainty that affect traditional absolute abundance analyses, including errors in oscillator strengths, limitations in 1D LTE atmospheric models, and systematic imperfections in continuum placement \citep[e.g.][]{Bedell2014}. Consequently, differential spectroscopy offers significantly higher internal precision and allows us to detect subtle star-to-star abundance variations within GCs.

For this analysis, we employed the Python-based tool \texttt{Q2} \citep{Ramirez2014}, which is specifically designed for differential abundance studies. \texttt{Q2} interfaces with \texttt{MOOG} \citep{Sneden1973}, a widely used 1D LTE radiative transfer code, and utilises a grid of $\alpha$-enhanced Kurucz model atmospheres \citep{Castelli2003IAUS..210P.A20C}. As input, \texttt{Q2} requires the equivalent widths (EWs) of spectral lines. To improve the robustness of the EW measurements, we measured EWs independently with two automated tools:
\texttt{REvIEW}\footnote{\url{https://github.com/madeleine-mckenzie/REvIEW/tree/main}} \citep{McKenzie2022}, designed for fitting EWs in high-resolution spectra, and \texttt{ARES}\footnote{\url{https://github.com/sousasag/ARES}} \citep{Sousa2007}, a widely used EW measurement code. We visually inspected all line fits and matched transitions line by line. We then compared the two sets of measurements using a weighted linear regression to identify systematic offsets and trends, and transformed the \texttt{REvIEW} EWs onto the \texttt{ARES} scale. It is worth noticing that, because \texttt{REvIEW} does not provide line-by-line EW error, we only used the error reported by \texttt{ARES} to perform the weighted regression between the \texttt{ARES} and \texttt{REvIEW} measurements and to derive the calibration relation between the two methods. After correcting for any systematic offset and trend, we adopted the final EW for each line as the average of the calibrated \texttt{REvIEW} measurement and the original \texttt{ARES} value. Finally, we rejected lines with EWs smaller than 5m\AA\ ~to avoid weak lines that are strongly affected by  noise and continuum placement and larger than 100m\AA ~to avoid saturated lines and lines where the difference between the two methods exceeded 20\% of the \texttt{ARES} EW, to remove clearly discrepant measurements.

\begin{figure*}
        \centering
        \includegraphics[width=\textwidth]{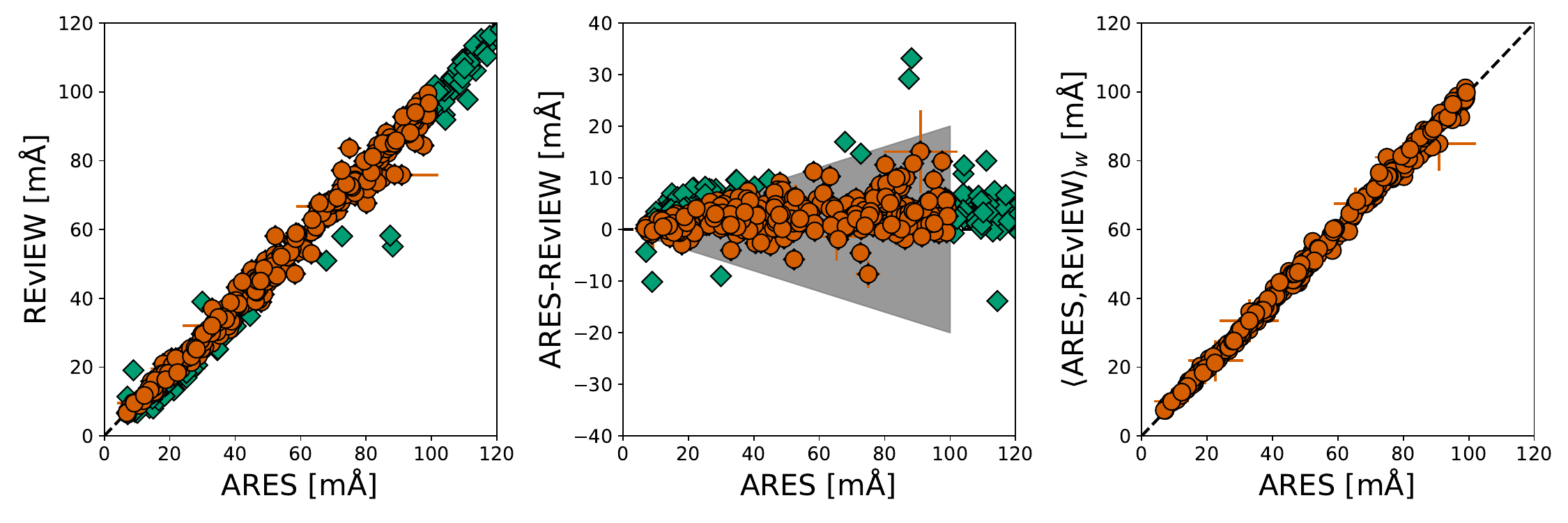}
        \caption{Comparison of EW measurements obtained with \texttt{ARES} and \texttt{REvIEW} for star ‘15590’ in NGC~6752. Green and orange symbols represent discarded and selected measurements, respectively. Left panel: Comparison of EWs measured with  \texttt{REvIEW} and \texttt{ARES}. The \texttt{REvIEW} EWs are plotted as a function of the \texttt{ARES} measurements. Middle panel: Difference between the EWs measured with the two methods as a function of the \texttt{ARES} EWs. Only measurements lying within the grey shaded region are included in the subsequent analysis. Right panel: Weighted average EWs derived from the two methods for lines with EWs between 5 and 100 m\AA.} 
        \label{fig:ReviewAres}
\end{figure*}

Figure~\ref{fig:ReviewAres} illustrates the comparison between the EWs derived from \texttt{REvIEW} and \texttt{ARES} and the difference between the two measurements in the left and middle panels, respectively, for the star \texttt{'15590'} of the GC NGC~6752. The results of both methods seem to show a slight trend and offset that was corrected by computing the weighted averaged EWs (right panel). In the middle panel, we also add a shaded grey area that indicates the EW limits that we imposed and discussed in the last paragraph.

For differential analysis, we grouped stars within each GC according to the effective temperature (\teff) reported by \citet{Carretta2009u} as a first guess, requiring that the maximum temperature difference within a group does not exceed 100 K. This threshold is considerably more stringent than those commonly adopted in the literature -- which are typically around 200–250 K  \citep[see e.g.][]{Monty2023} -- and plays a key role in improving our sensitivity to subtle abundance variations and in mitigating potential trends of differential abundances with stellar parameters. To ensure statistical robustness, we only retained groups that contained at least four stars. Within each group, we selected as a reference the star whose input \teff lay closest to the group mean, irrespective of whether it belonged to the FG or SG. We then used this star as the baseline for the differential comparison.

\begin{figure*}
        \centering
        \includegraphics[width=0.9\textwidth]{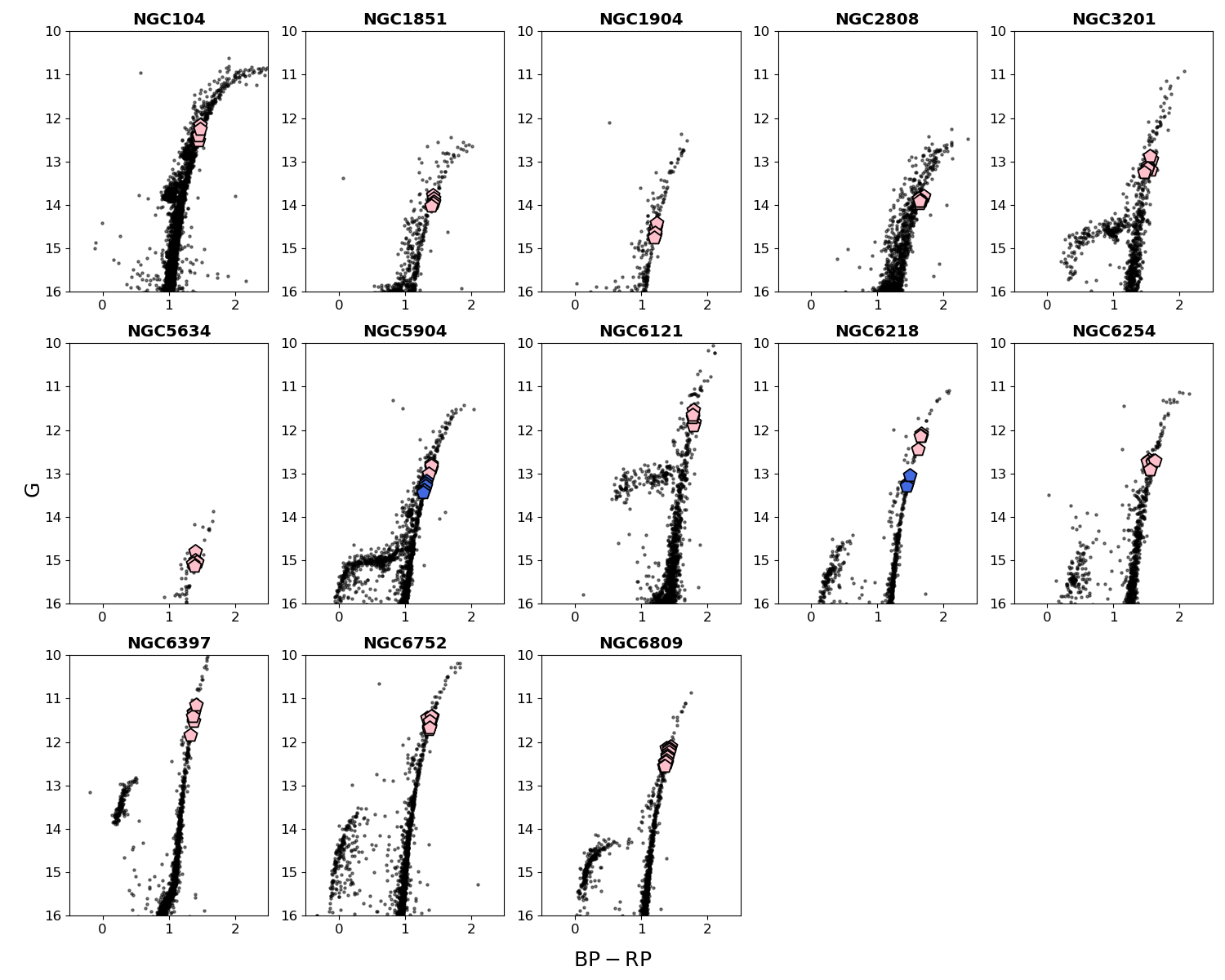}
        \caption{Colour–magnitude diagrams of the 13 GCs analysed using Gaia DR3 photometry. Small black symbols indicate cluster members based on the selection done by \citet{Vasiliev2021}, while large pentagons mark the stars in our sample. For clusters with more than one subsample (NGC~5904 and NGC~6218), the pentagons appear in two different colours to distinguish the subsamples.}
        \label{fig:cmds}
\end{figure*}

Figure~\ref{fig:cmds} shows the colour-magnitude diagrams for each GC in our sample, constructed using {\em Gaia} DR3 photometry \citep{GaiaDR3_2023}. For the purposes of our analysis, we further divided the groups NGC~5904 and NGC~6218 into two distinct subsamples, shown in different colours, to maximise internal homogeneity within each group.

Finally, we defined the differential abundances of a given line as $\delta A_{i} = A_{i}^{star}-A_{i}^{reference}$. Then the mean Fe differential abundance of a star with respect to the reference is given by
$$\langle \delta A_{i}^{Fe} \rangle = \frac{1}{N} \sum_{i=1}^{N} \delta A_{i}^{Fe}= \Delta^{Fe},$$ where $N$ is the number of lines measured. Accordingly, the standard deviation is given by 
$$\sigma^{Fe}=\sqrt{\sum^{N}_{i=i} \frac{(\delta A_{i}-\Delta^{Fe})^2}{N}}.$$

Therefore, the mean differential abundance of the group ($\Delta^{Fe}_g$) is given by the average $\Delta^{Fe}$ of the stars that belong to the group. We estimated the uncertainty on the observed Fe abundance spread through Monte Carlo error propagation. For each star, we perturbed the measured abundance by drawing random values from a Gaussian distribution centred on the observed abundance, with a standard deviation equal to the individual abundance uncertainty. For each cluster, we generated 10,000 synthetic realisations and calculated the Fe abundance dispersion for each realisation. We used the standard deviation of the resulting distribution of simulated spreads as the uncertainty on the observed spread. This approach accounts for the individual stellar abundance uncertainties and the finite size of the stellar sample.

\subsection{Determination of stellar parameters and Fe abundances}

Stellar parameters and iron abundances were re-derived using a differential approach implemented in \texttt{Q2} \citep{Ramirez2014}. For each GC, a reference star was selected, and its absolute parameters and iron abundance were first determined within \texttt{Q2}. These values were then used as a baseline for calculating the differential abundances of the remaining stars within the same group.

As initial estimates of the stellar parameters, we adopted the values reported by \citet{Carretta2009u} and \citet{Carretta2017}, which provide a homogeneous and well-tested reference. Using these parameters, we derived preliminary iron abundances and excluded spectral lines that deviated by more than 3$\sigma$ from the mean abundance. We then refined the stellar parameters iteratively, adopting the literature values as starting guesses.

Since \texttt{Q2} applies a differential line-by-line method, the optimisation of stellar parameters is achieved by minimising (i) the slope of Fe I abundances as a function of excitation potential, (ii) the slope of Fe I abundances against reduced EW, and (iii) the discrepancy between Fe I and Fe II abundances. As a result, the final parameters represent the set of differential values that best satisfy excitation and ionisation equilibrium relative to the chosen reference star, rather than yielding absolute measurements for individual stars. It is worth noticing that the final set of stellar parameters does not affect our conclusions, as our analysis focuses on differential abundances rather than absolute parameter values, and therefore all quantities related to the stellar parameters are effectively subtracted. Moreover, the stellar parameters adopted from \citetalias{Carretta2009u} are generally very close to those independently re-derived in this work, ensuring internal consistency within the analysis. Table \ref{tab:stellar_parameters} lists the star's IDs \citep[taken from][]{Carretta2009u} and the resulting differential stellar parameters for all stars included in our analysis. We indicate the reference stars with the suffix ‘-r’. Our final sample contains 92 sibling stars, organised into 15 groups across 13 GCs.

\begin{table*}[!ht]
    \centering
   \caption{IDs and their corresponding RA and DEC coordinates, differential stellar parameters, along with the respective errors of our sample of stars in each GC.}
    \begin{tabular}{ccccccccc}
    \hline \hline
       Cluster & ID  & RA & DEC &\teff & $\log$ g & [Fe/H] & $\nu_t$ & $\Delta^{Fe}$\\ \hline
       NGC~104 & 13795-r & 6.30535417 & -72.13067500 & 4332$\pm$37 & 1.81$\pm$0.17 & -0.71$\pm$0.04 & 1.84$\pm$0.09 &  0.00$\pm$0.00\\
       NGC~104 & 14583   & 5.63324583 & -72.11360833 & 4316$\pm$19 & 1.92$\pm$0.12 & -0.74$\pm$0.03 & 1.65$\pm$0.07 & -0.02$\pm$0.03\\
       NGC~104 & 18623   & 5.83738333 & -72.08840278 & 4324$\pm$20 & 1.92$\pm$0.12 & -0.77$\pm$0.03 & 1.79$\pm$0.07 & -0.04$\pm$0.03\\
       NGC~104 & 20002   & 5.87197500 & -72.08006944 & 4263$\pm$27 & 1.79$\pm$0.13 & -0.75$\pm$0.04 & 1.75$\pm$0.07 & -0.05$\pm$0.03\\
       NGC~1851& 14080-r & 78.4111875 & -40.12765278 & 4458$\pm$45 & 1.67$\pm$0.17 & -1.06$\pm$0.04 & 1.76$\pm$0.09 &  0.00$\pm$0.00\\
        ...     & ...  & ...  & ...   & ...  & ... \\ \hline \hline  
    \end{tabular}
    \tablefoot{The reference star is indicated with the suffix ‘-r’. The full table is available in electronic form at the CDS.}
    \label{tab:stellar_parameters}
\end{table*}

\subsection{Robustness of the method}
\label{Sec: CompLite}

The differential abundance method has been applied only sparingly in the literature to the study of GCs \citep[see e.g.][]{McKenzie2022,Yong2013,Lardo2023,Monty2023}. In a recent work, \citet{McKenzie2022} used this technique for six stars in the GC NGC~6656 (M~22), obtaining high-precision measurements for 18 chemical elements based on UVES spectra. To verify the robustness of our methodology, we retrieved spectra for a subset of their target stars and re-analysed them using our differential approach. For this purpose, we selected the stars III-3, III-14, and IV-102, which, according to their analysis, differ in effective temperature by less than 100 K.

\begin{table}[!ht]
    \centering
   \caption{Comparison of stellar parameters obtained using our method and the ones reported by \citet{McKenzie2022} for the three sibling stars in NGC~6656.}
   \resizebox{\columnwidth}{!}{
\begin{tabular}{ccccccccc}
\hline \hline
       & \multicolumn{4}{c}{This work} & \multicolumn{4}{c}{Mckenzie+2022} \\ \hline
ID     & \teff  & $\log$ g  & [Fe/H]    & $\nu_t$   & \teff   & $\log$ g   & [Fe/H]     & $\nu_t$    \\ \hline
III-3  & 4066  & 0.52  & -1.68  & 2.18 & 4041   & 0.25   & -1.78   & 2.29  \\
III-14 & 4061  & 0.27  & -1.78  & 2.47 & 4038   & 0.12   & -1.87   & 2.24  \\
IV-102 & 4081  & 0.24  & -1.86  & 2.55 & 4043   & 0.10   & -1.97   & 2.43  \\ \hline \hline
\end{tabular}}
    \label{tab:comp_stellar_parameters}
\end{table}

\begin{figure*}
        \centering
        \includegraphics[width=0.95\textwidth]{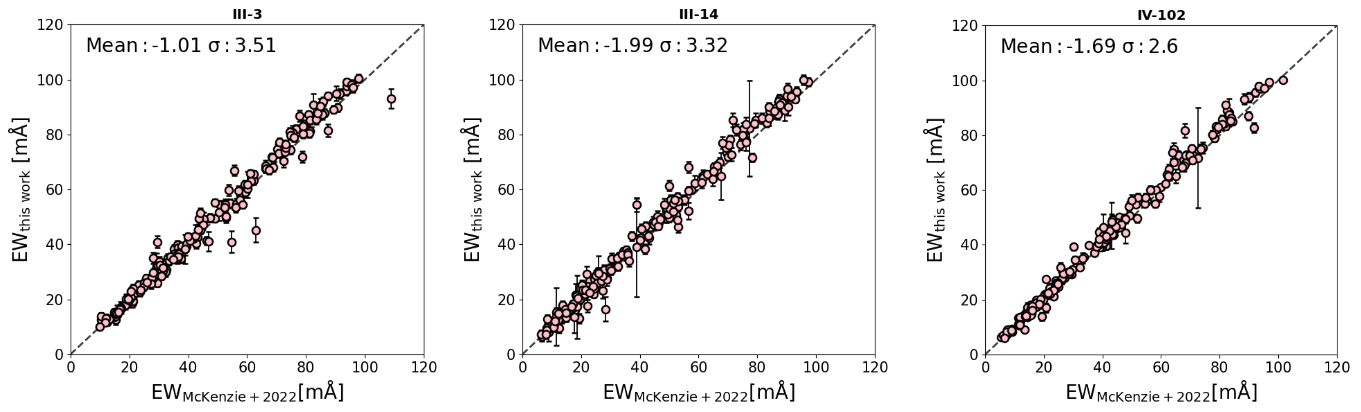}
        \caption{Comparison between the EWs measured by \citet{McKenzie2022} and those obtained with our method for the M~22 stars III-3, III-14, and IV-102. Each panel reports the mean difference and standard deviation.}
        \label{fig:comp_mii-14}
\end{figure*}

Figure~\ref{fig:comp_mii-14} compares the EWs measured in our study with those reported by \citet{McKenzie2022} for these stars. The two sets of measurements show good agreement, with mean offsets of approximately -1.01, -1.99, and -1.69 m\AA~ and standard deviations of 3.51, 3.32, and 2.60 m\AA~ for III-3, III-14, and IV-102, respectively. This level of consistency supports the reliability of our EW determinations and the robustness of our method. Table~\ref{tab:comp_stellar_parameters} lists the stellar parameters derived using our approach alongside those reported by \citet{McKenzie2022}. Using star III-3 as the reference, we derived average differential [Fe/H] values of –0.10 and –0.18 dex for III-14 and IV-102, respectively, while \citet{McKenzie2022} reported corresponding values of -0.09 and -0.19 dex. The close agreement between the two analyses further confirms the validity of our results.

\section{Results}
\label{Sec: Results}

\subsection{Internal iron spread by cluster}

Before assessing the statistical significance of the inferred iron variations, we first examine the observed differential Fe spreads measured in each GC. Across the sample, the absolute values of the dispersions remain small. In most clusters, the measured spreads lie within the range $\sigma^{Fe}_g \sim$ 0.02–0.05 dex, smaller than those reported in previous photometric studies (e.g. \citealt{Legnardi2022,Latour2025}). These results do not provide support for the existence of a spread much beyond the current detection limits of high-resolution spectroscopy.

The limited number of stars analysed in each cluster complicates the interpretation of these spreads, as several samples contain fewer than ten stars. Consequently, the observed dispersions alone do not allow for a robust assessment of statistical significance. To address this limitation, we complemented the observational analysis with a Monte Carlo approach designed to determine whether the measured spreads can arise solely from the reported uncertainties.

To test whether the internal Fe spread observed in each GC can be explained entirely by measurement errors, we compared the observed dispersion of each group ($\sigma^{Fe}_g$) with a simulated spread. We generated the simulated distributions through 10,000 Monte Carlo iterations, assuming a normal distribution centred on the mean differential abundance of the group ($\Delta^{Fe}_g$), with the spread determined only by the uncertainties reported by \texttt{Q2}.

\begin{figure*}
     \centering
     \begin{subfigure}{0.285\textwidth}
         \includegraphics[width=\linewidth]{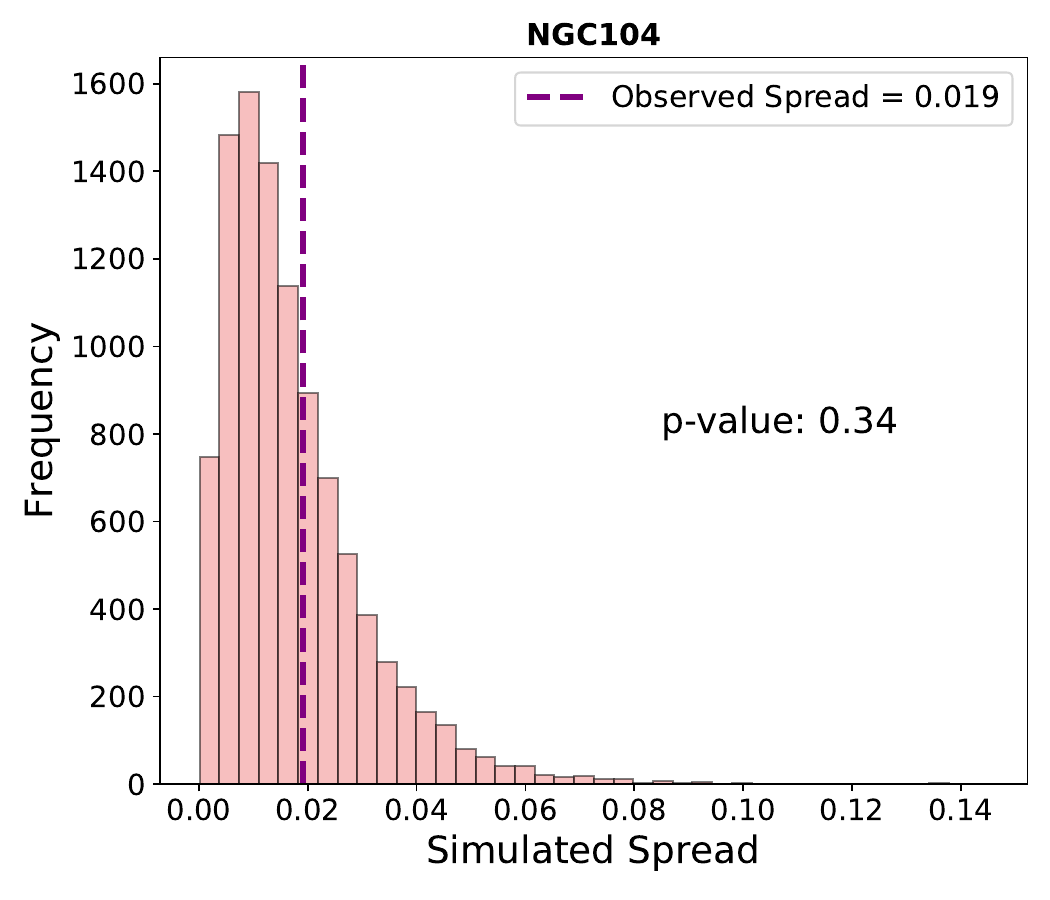}
     \end{subfigure}
     \hfill
     \begin{subfigure}{0.285\textwidth}
         \includegraphics[width=\linewidth]{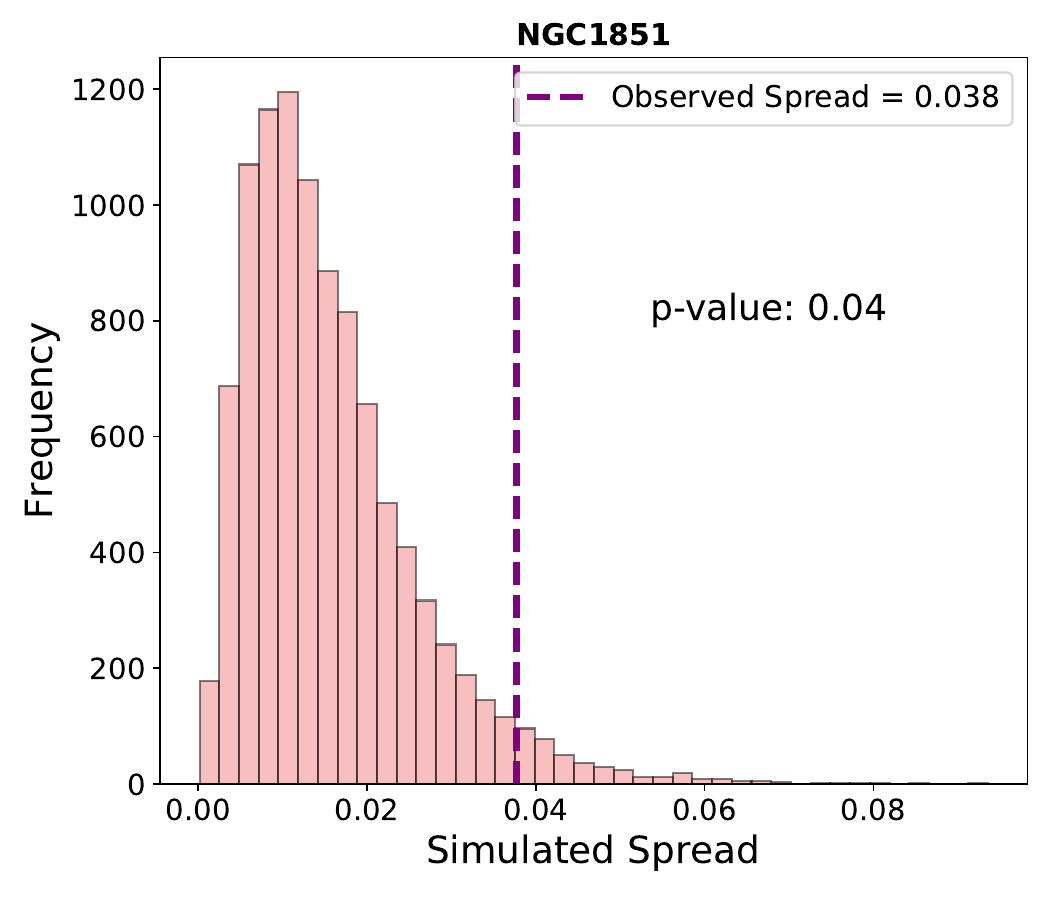}
     \end{subfigure}
     \hfill
     \begin{subfigure}{0.285\textwidth}
         \includegraphics[width=\linewidth]{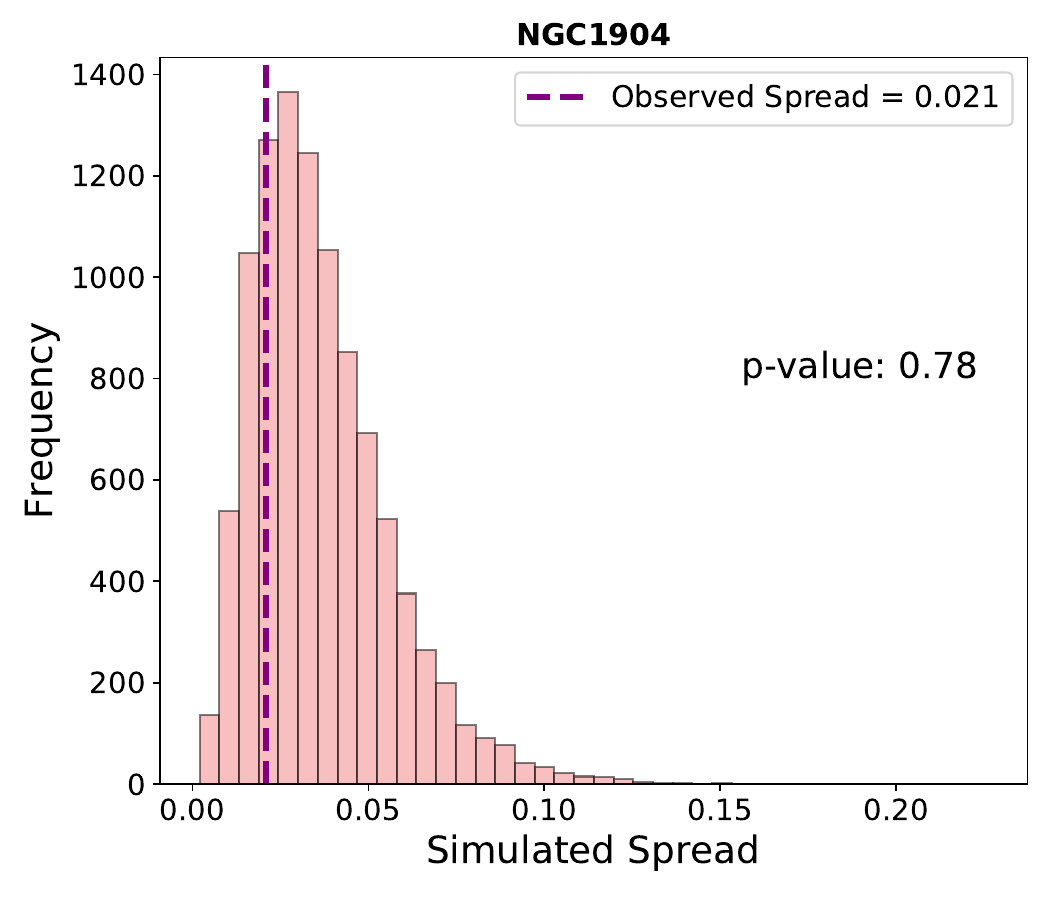}
     \end{subfigure}
     \begin{subfigure}{0.285\textwidth}
         \includegraphics[width=\linewidth]{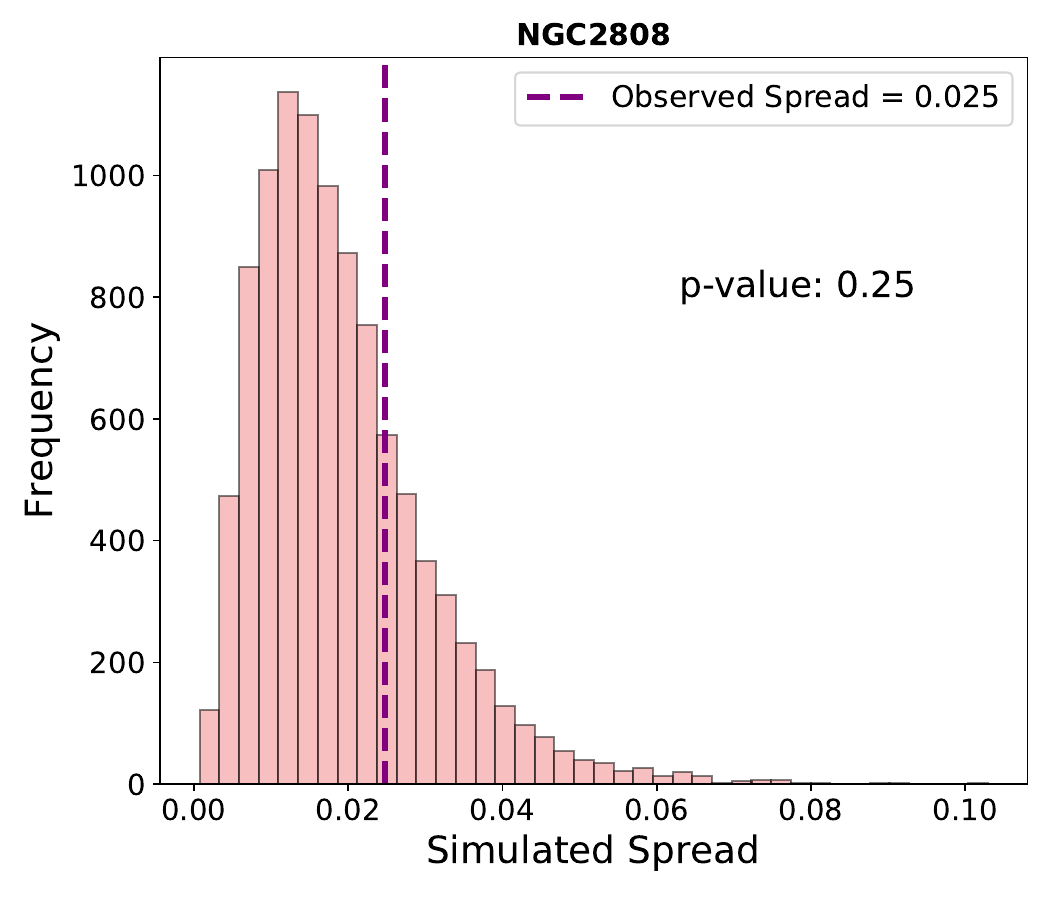}
     \end{subfigure}
     \hfill
     \begin{subfigure}{0.285\textwidth}
         \includegraphics[width=\linewidth]{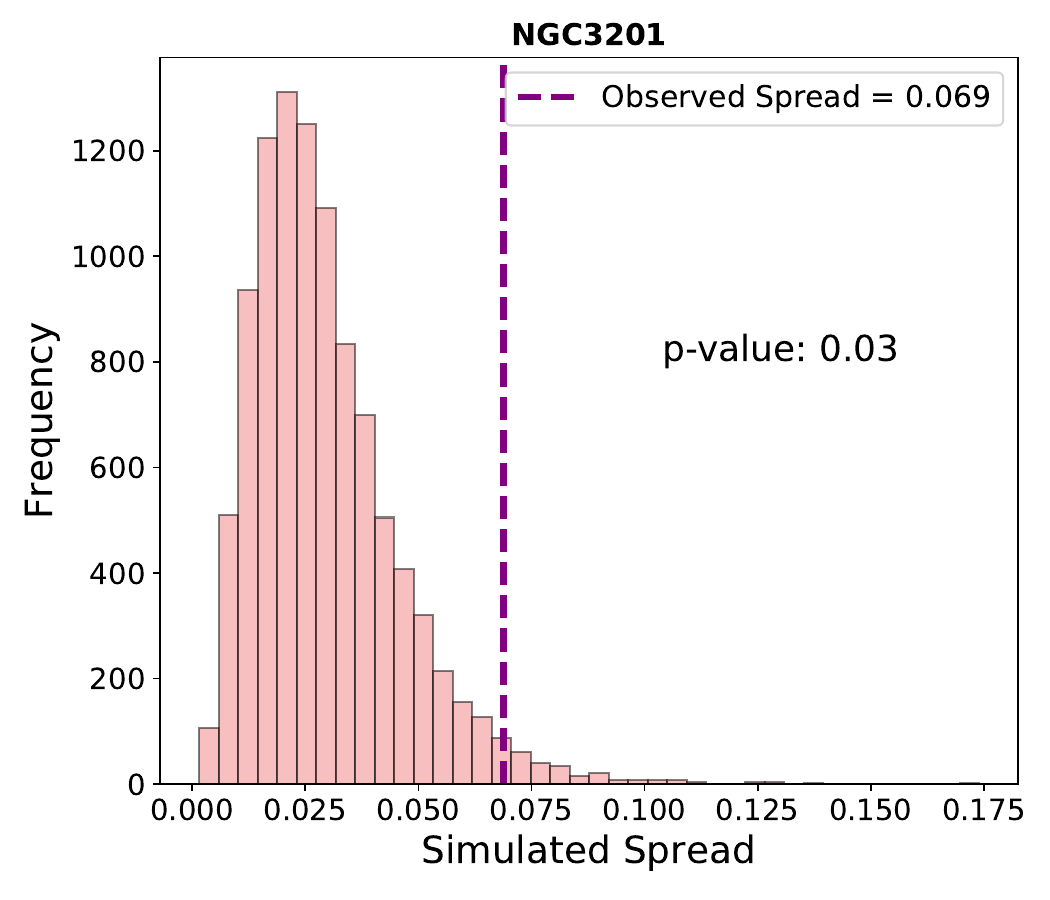}
     \end{subfigure}
     \hfill
     \begin{subfigure}{0.285\textwidth}
         \includegraphics[width=\linewidth]{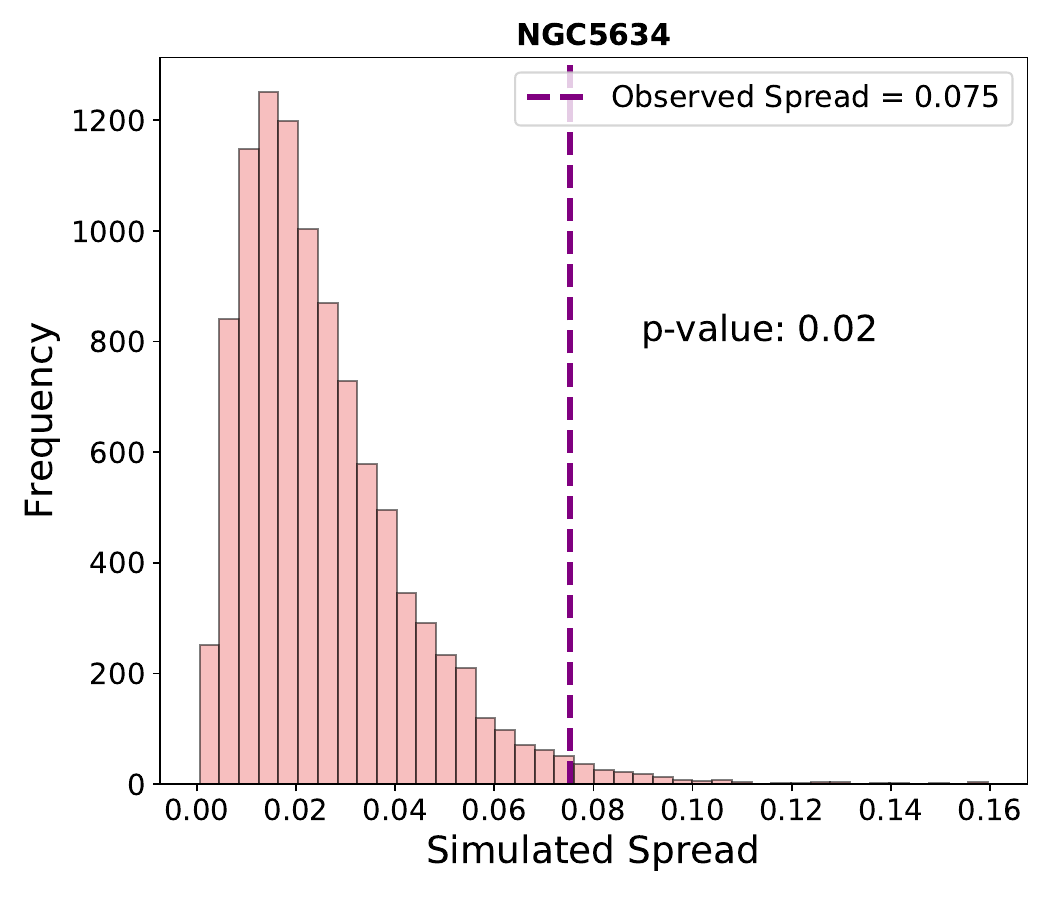}
     \end{subfigure}
     \hfill
     \begin{subfigure}{0.285\textwidth}
         \includegraphics[width=\linewidth]{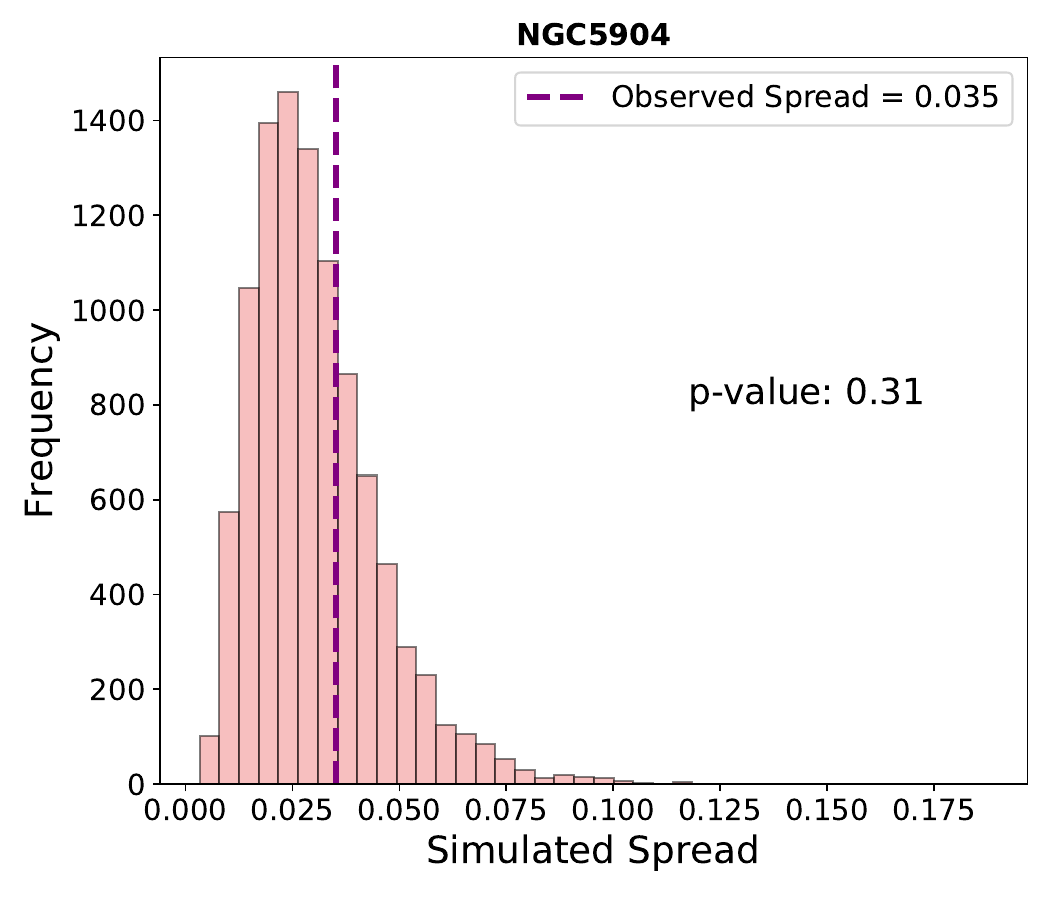}
     \end{subfigure}
     \hfill
     \begin{subfigure}{0.285\textwidth}
         \includegraphics[width=\linewidth]{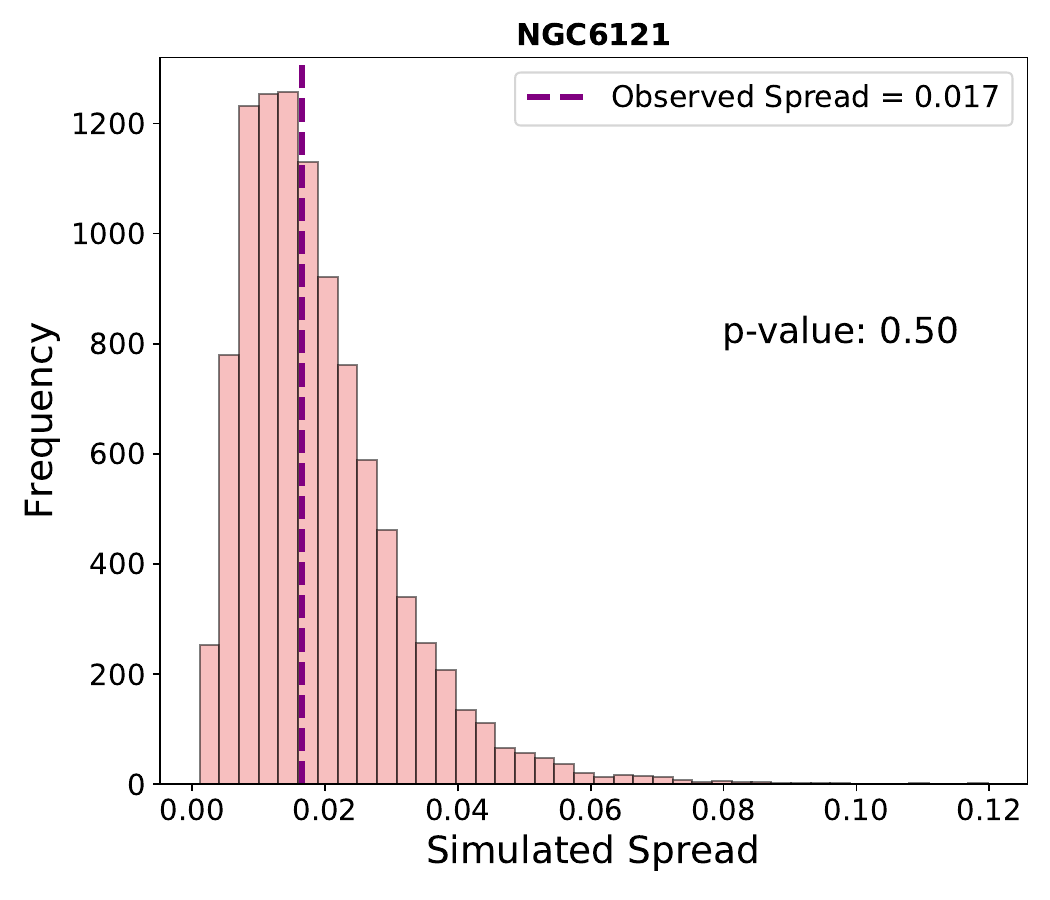}
     \end{subfigure}
     \hfill
     \begin{subfigure}{0.285\textwidth}
         \includegraphics[width=\linewidth]{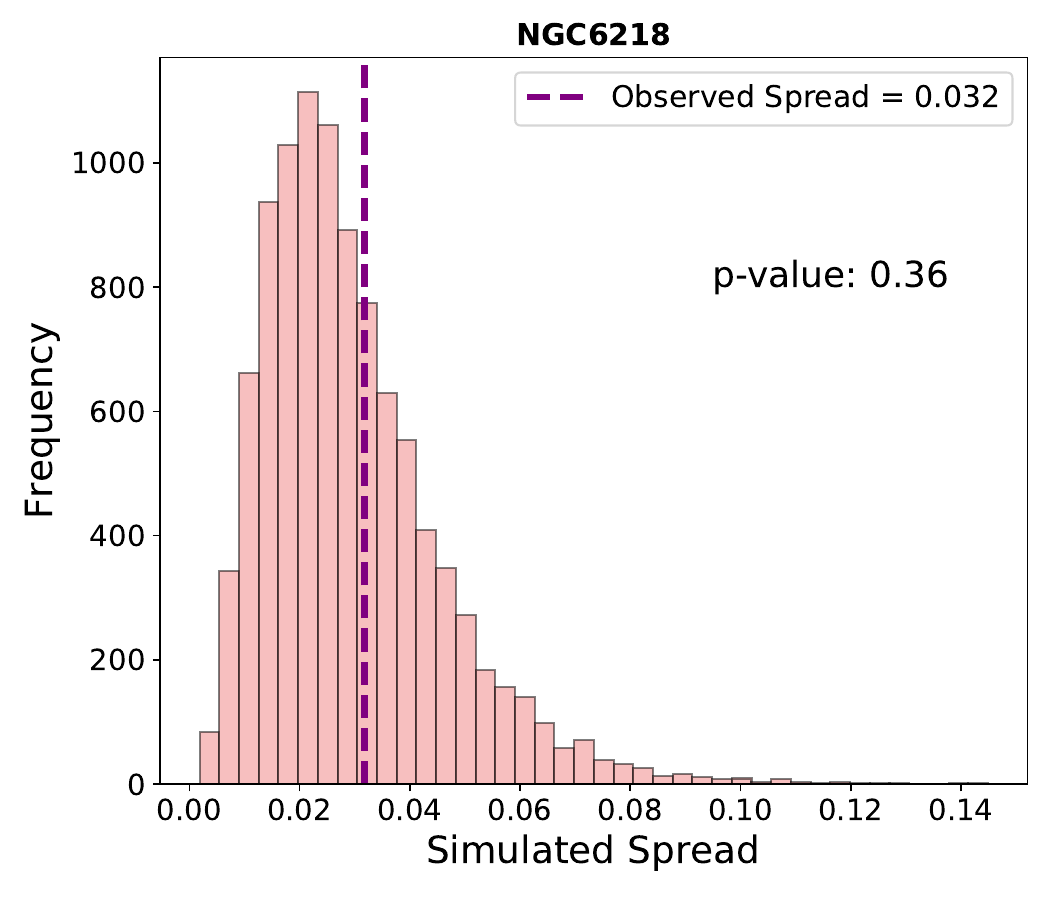}
     \end{subfigure}
     \hfill
     \begin{subfigure}{0.24\textwidth}
         \includegraphics[width=\linewidth]{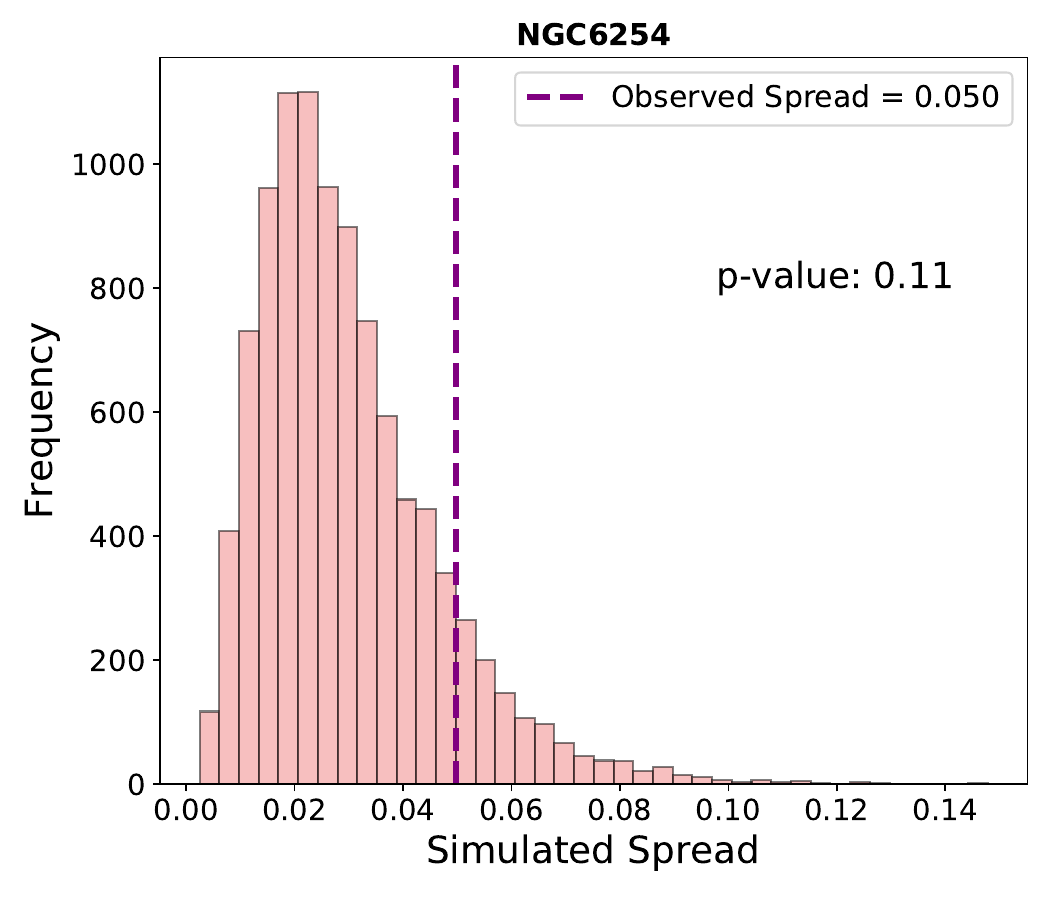}
     \end{subfigure}
     \hfill
     \begin{subfigure}{0.24\textwidth}
         \includegraphics[width=\linewidth]{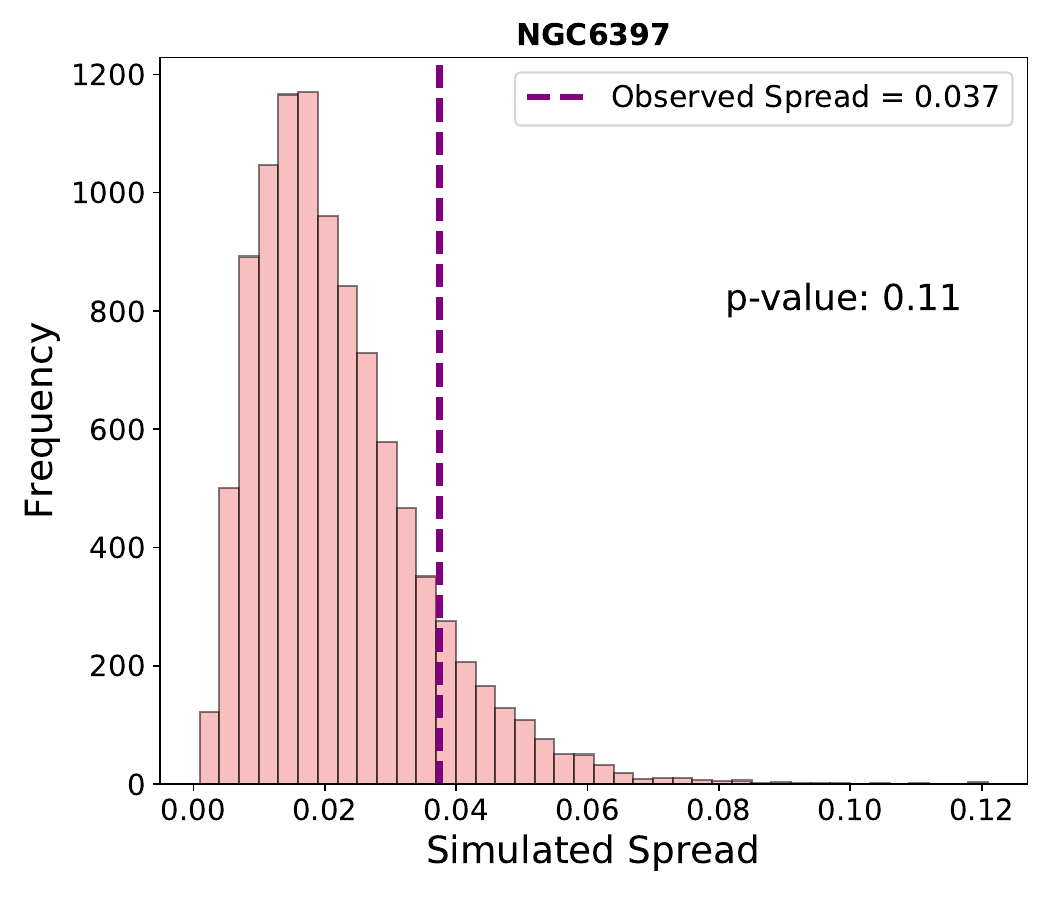}
     \end{subfigure}
     \hfill
     \begin{subfigure}{0.24\textwidth}
         \includegraphics[width=\linewidth]{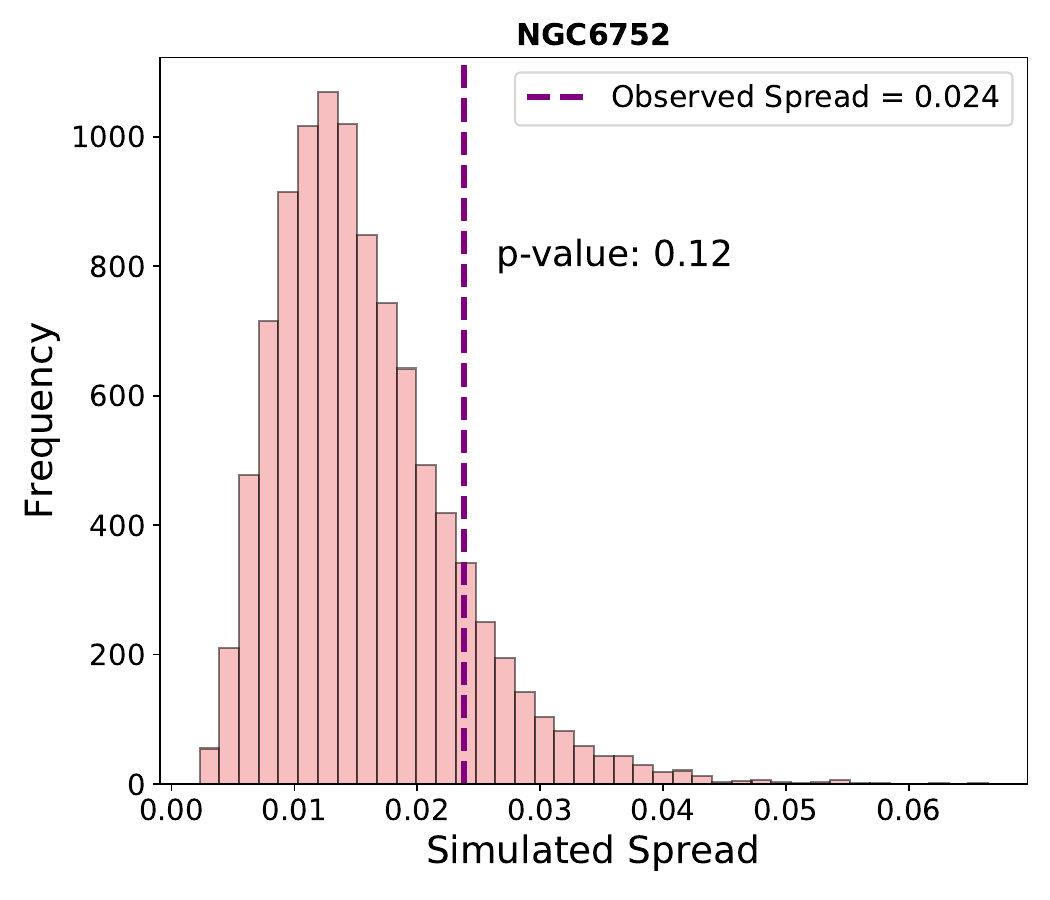}
     \end{subfigure}
    \hfill
     \begin{subfigure}{0.24\textwidth}
         \includegraphics[width=\linewidth]{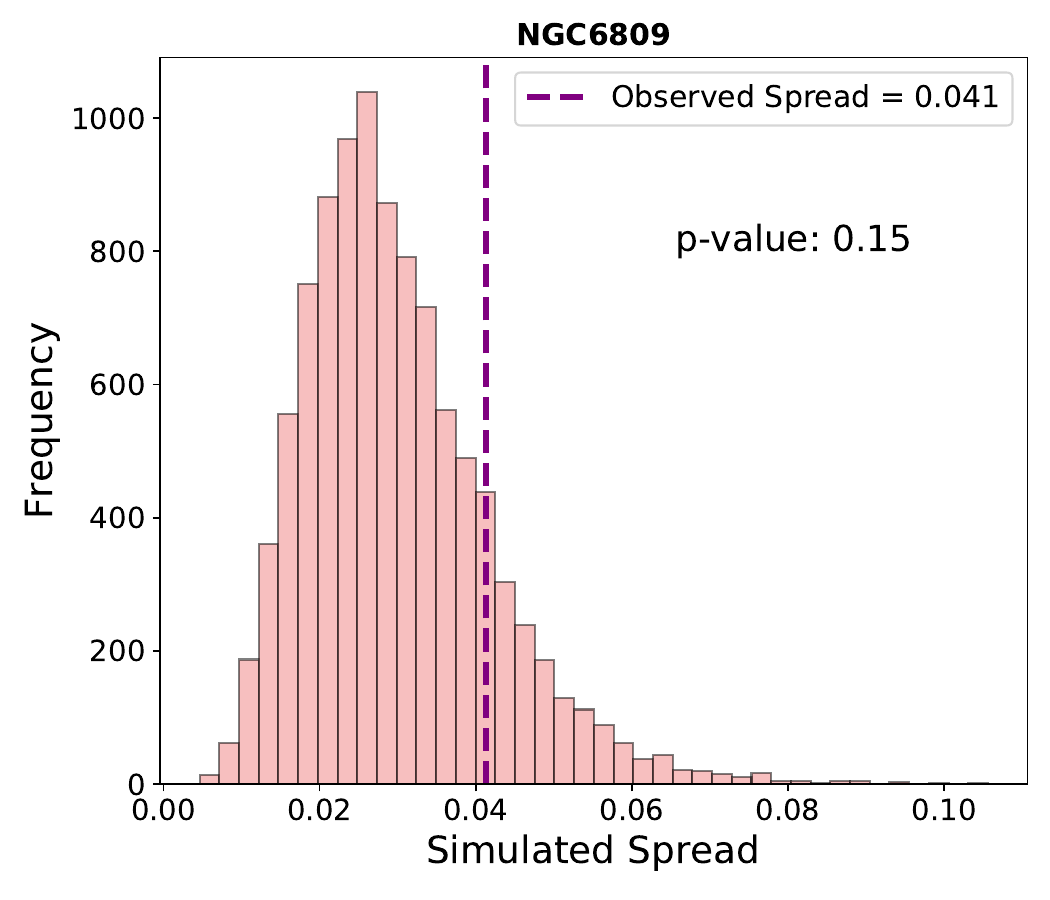}
     \end{subfigure}
     \caption{Histogram of 10,000 Monte Carlo simulations of Fe spread. The dashed purple line indicates the observed Fe spread in our sample of GCs.}
     \label{fig:FeI_histograms_a}
\end{figure*}

Figure~\ref{fig:FeI_histograms_a} presents the results for the full set of GCs, showing histograms of the simulated spreads obtained from the 10,000 iterations. The dashed purple line marks the observed spread for each cluster ($\sigma^{Fe}_g$). We evaluated the statistical significance of the observed values using the p value, defined as the fraction of simulations in which the simulated spread exceeds the observed one. In this work, we classify a dispersion as significant when the p value is $\leq 0.05$, and as not significant when it exceeds this threshold.

We grouped the clusters according to the statistical significance of their Fe spread. The clusters NGC~104, NGC~1904, NGC~2808, NGC~5904, NGC~6121, NGC~6218, NGC~6254, NGC~6397, NGC~6752, and NGC~6809 do not show significant Fe spreads, indicating that the observed dispersions in these systems can be fully explained by measurement uncertainties. In contrast, NGC~1851, NGC~3201, and NGC~5634 exhibit significant Fe spreads that cannot be attributed solely to measurement errors.

It should be noted that we identified two subsamples in each of NGC~5904 and NGC~6218. We initially defined these groups separately because their stellar temperatures differ by about 150 K in NGC~5904 and 250 K in NGC~6218. Nevertheless, under the assumption that the iron abundance remains chemically homogeneous within each cluster, we then merged the subsamples into single groups that comprise nine stars in NGC~5904 and seven stars in NGC~6218. The resulting p values -- approximately 0.31 for NGC~5904 and 0.36 for NGC~6218 -- indicate that the observed dispersions remain consistent with the measurement uncertainties reported and do not require intrinsic Fe variations. We also verified that our results remain stable, within the uncertainties, regardless of whether we select the reference star from one subsample or the other within each cluster.

\subsection{Internal iron spreads by generation}

We applied the same test to each GC separately for each stellar generation. To assign stars to a given generation, we adopted the classification of \citet{Carretta2009u}, which relies on [Na/Fe] abundances. Only five of the thirteen GCs in our sample include at least two stars per stellar generation. Most of the remaining clusters contain only one star in one of the generations. In particular, NGC~104, NGC~1851, NGC~1904, NGC~6121, NGC~6218, and NGC~6254 each include only one FG star, while our samples of NGC~2808 and NGC~5634 contain only one SG star.

\begin{figure}
        \centering
        \includegraphics[width=\columnwidth]{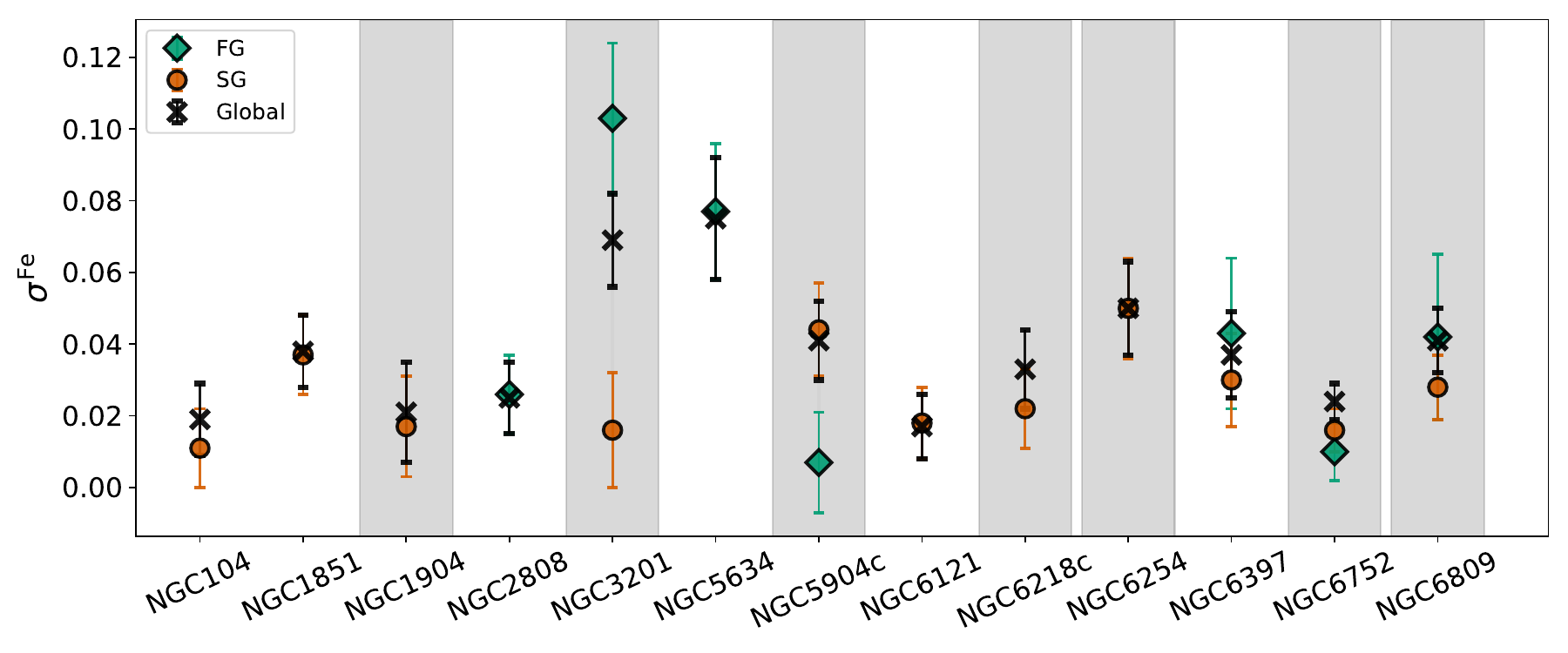}
        \caption{Measured abundance spreads for the full GC sample. Green and orange symbols denote the spreads of FG and SG stars, respectively, following the classification of \citet{Carretta2009u}. Black crosses indicate the spread measured across all cluster members. The shaded grey region highlights clusters with samples larger than six stars.}
        \label{fig:graph_sigma}
\end{figure}

Table~\ref{tab:Stats} summarises the observed abundance spreads for the full sample ($\sigma^{Fe}_g$) in each cluster, together with the spreads measured within each stellar generation. We denote the observed spreads among FG and SG stars as $\sigma^{Fe}_{FG}$ and $\sigma^{Fe}_{SG}$, respectively. The table also lists the number of stars assigned to each generation. Figure~\ref{fig:graph_sigma} presents the same information graphically: blue and orange circles represent the standard deviations of FG and SG stars ($\sigma^{Fe}_{FG}$ and $\sigma^{Fe}_{SG}$), while the black cross marks the differential standard deviation of the entire cluster ($\sigma^{Fe}_{g}$). The shaded green region highlights clusters with at least five analysed member stars.

\begin{figure*}
     \centering
     \begin{subfigure}{0.285\textwidth}
         \includegraphics[width=\linewidth]{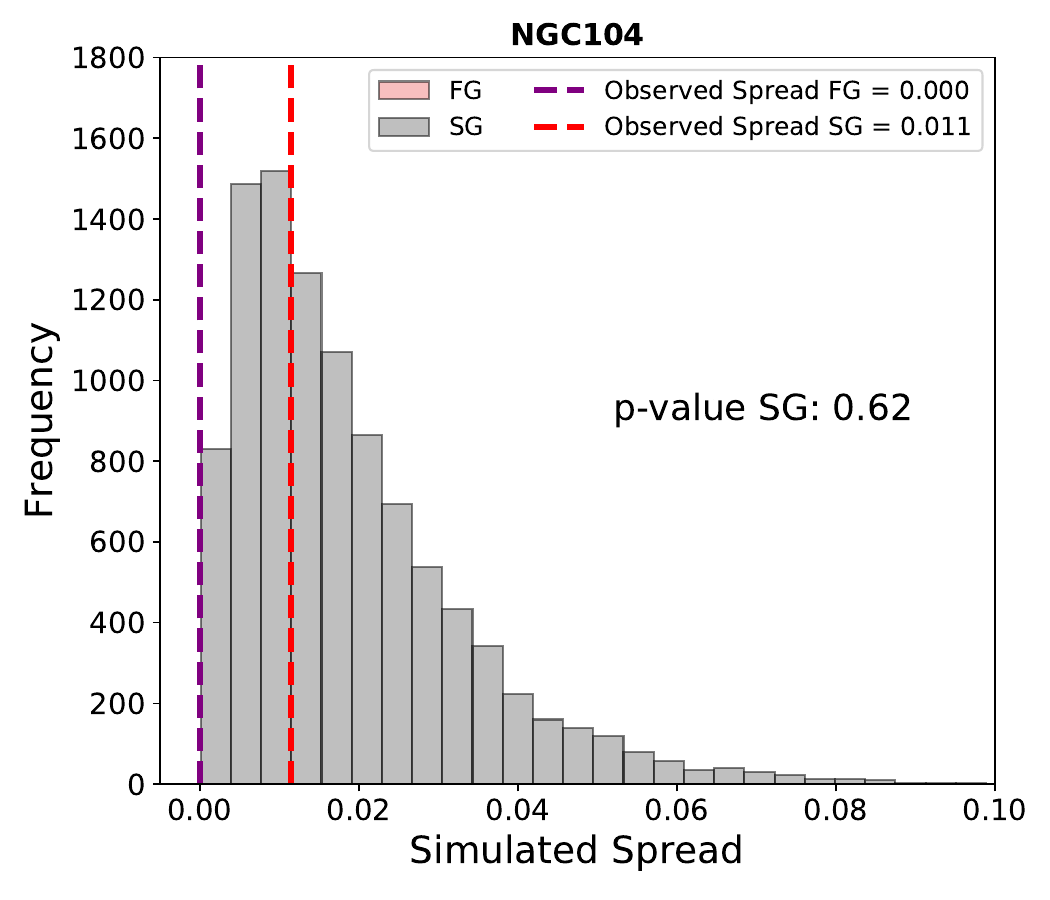}
     \end{subfigure}
     \hfill
     \begin{subfigure}{0.285\textwidth}
         \includegraphics[width=\linewidth]{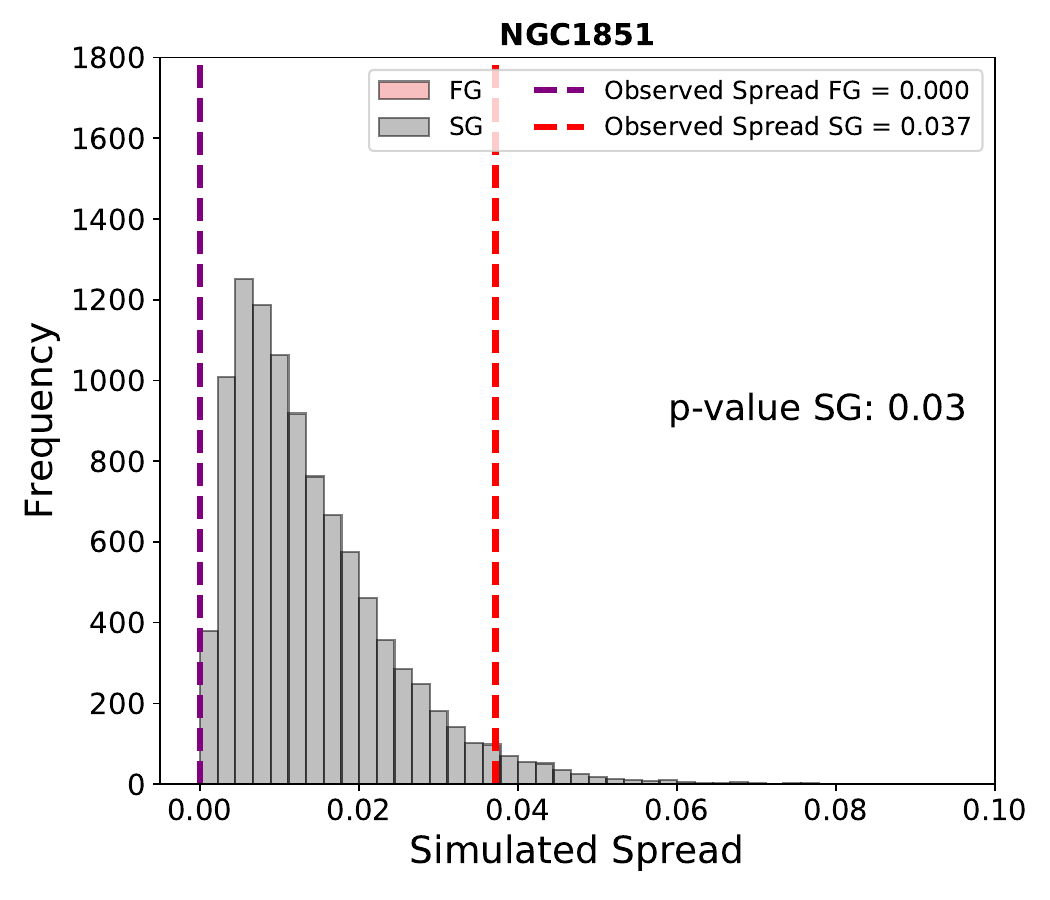}
     \end{subfigure}
     \hfill
     \begin{subfigure}{0.285\textwidth}
         \includegraphics[width=\linewidth]{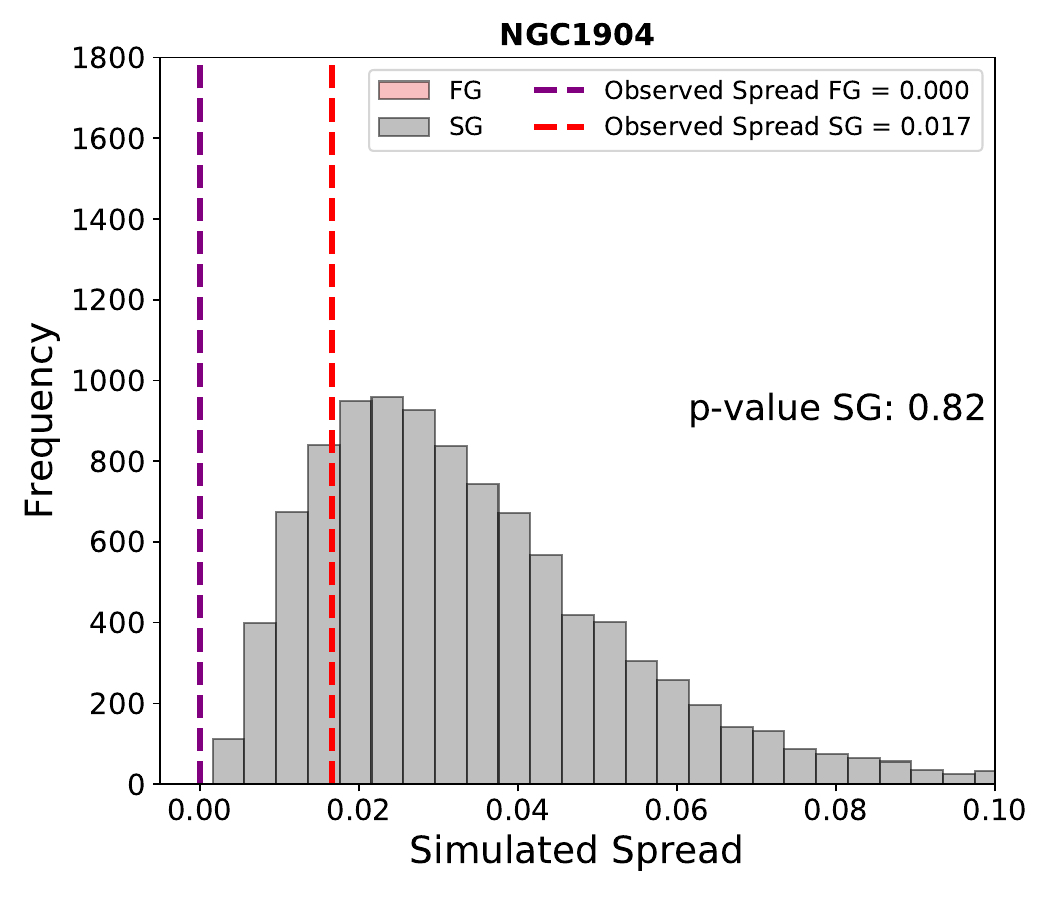}
     \end{subfigure}
     \hfill
     \begin{subfigure}{0.285\textwidth}
         \includegraphics[width=\linewidth]{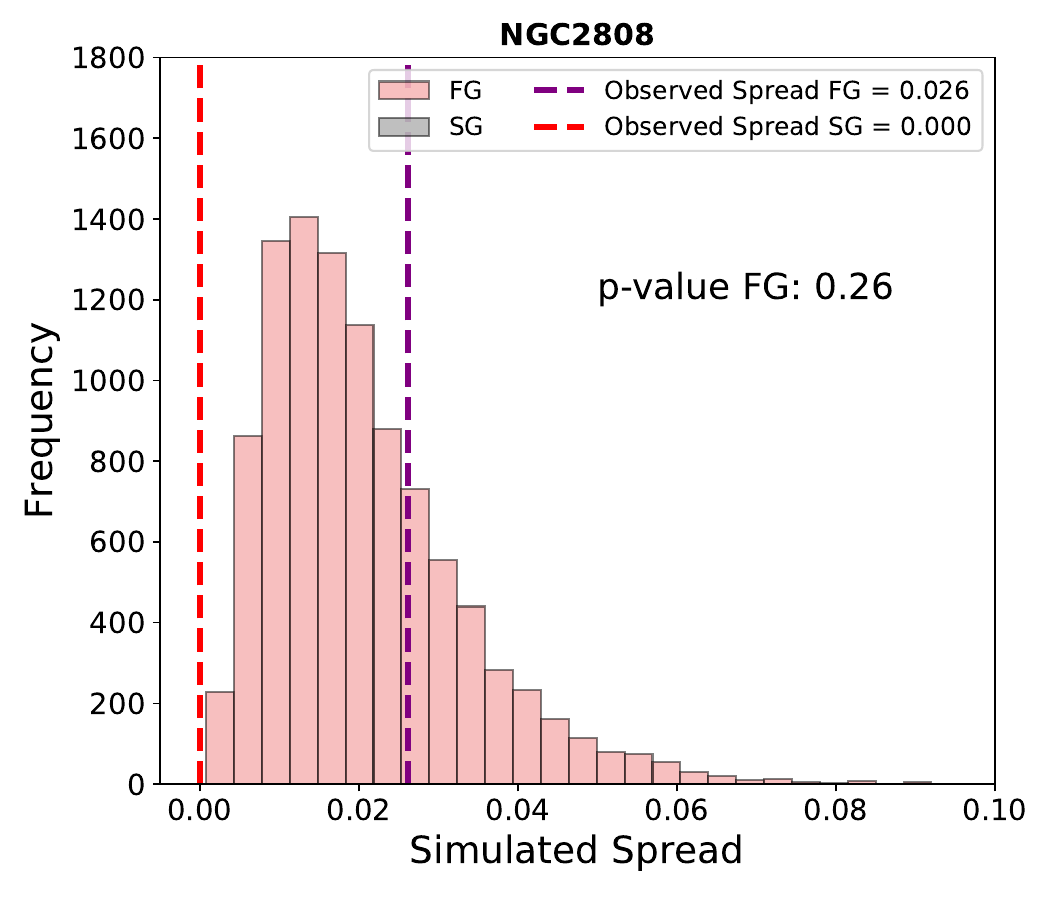}
     \end{subfigure}
     \hfill
     \begin{subfigure}{0.285\textwidth}
         \includegraphics[width=\linewidth]{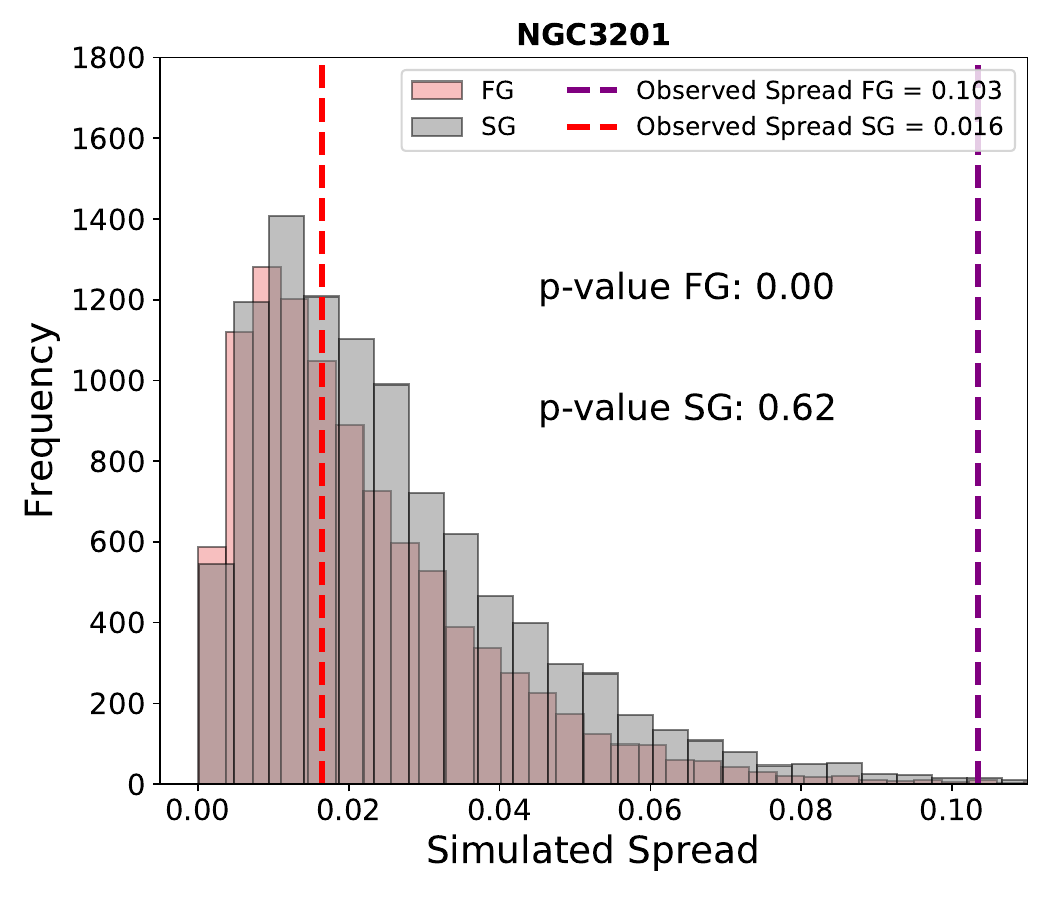}
     \end{subfigure}
     \hfill
     \begin{subfigure}{0.285\textwidth}
         \includegraphics[width=\linewidth]{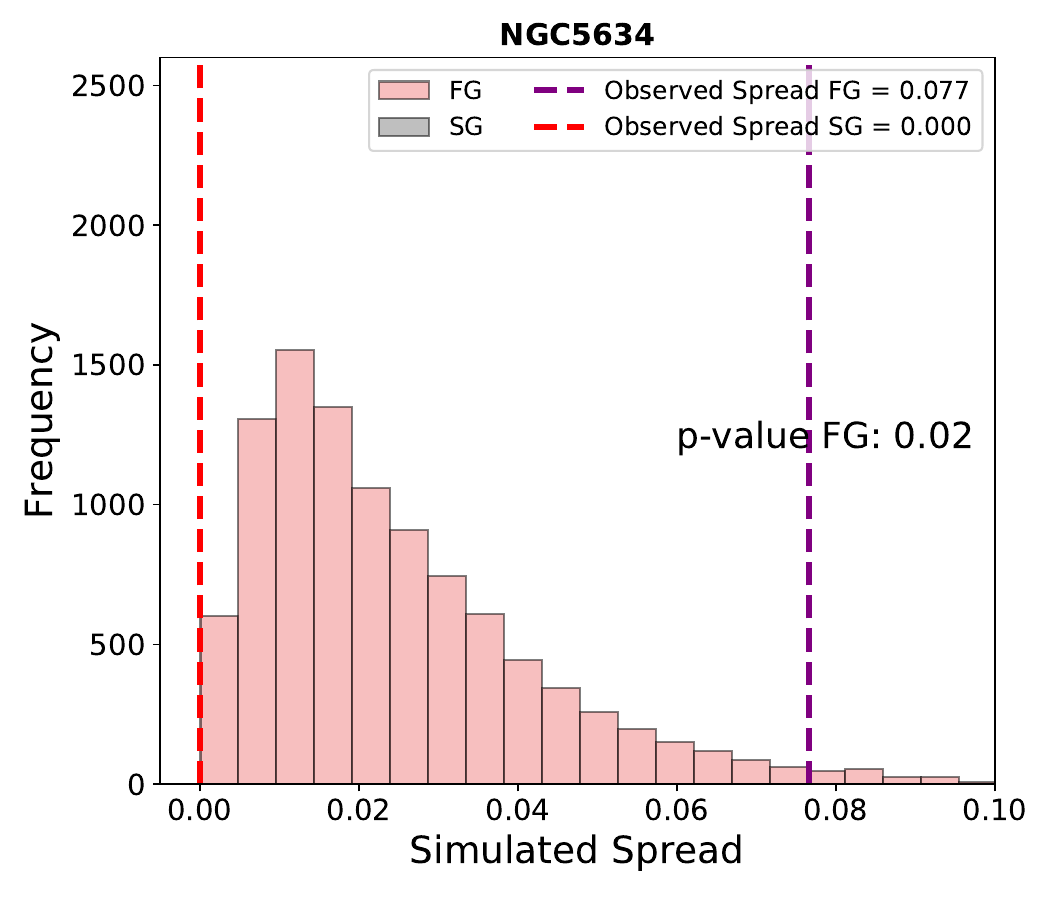}
     \end{subfigure}
     \hfill
     \begin{subfigure}{0.285\textwidth}
         \includegraphics[width=\linewidth]{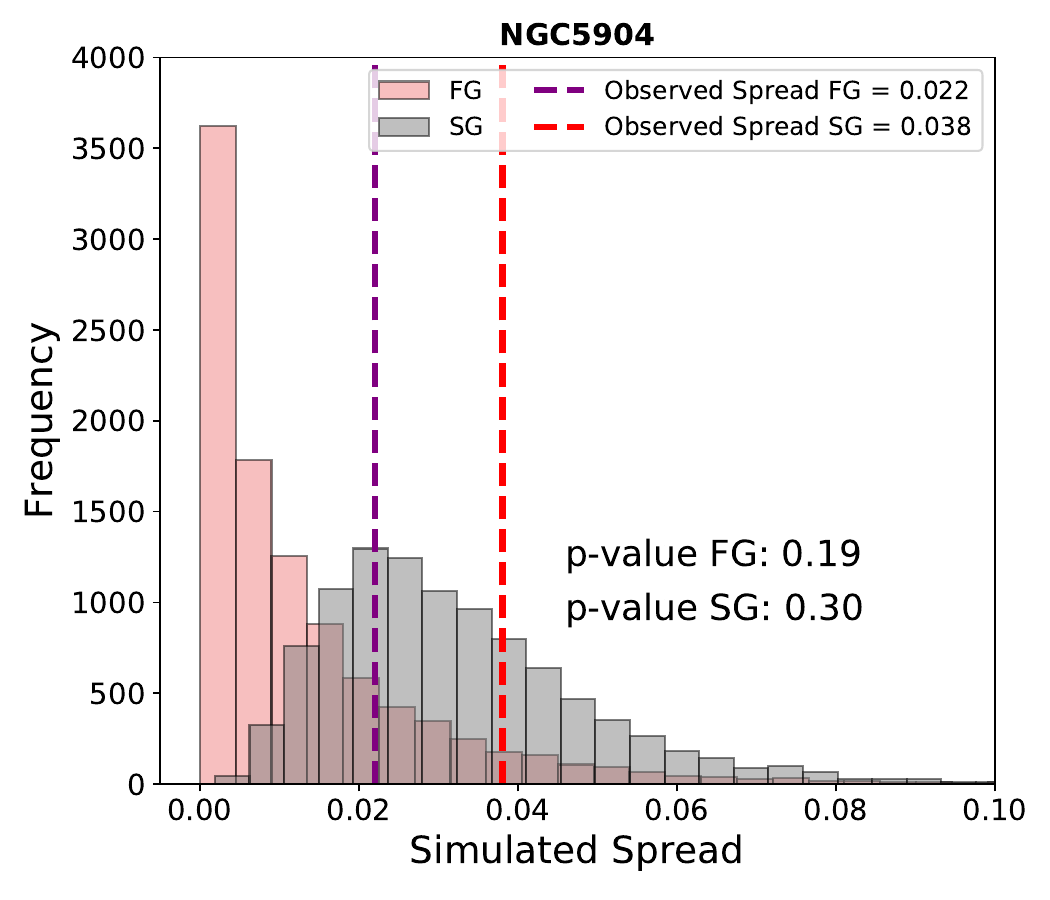}
     \end{subfigure}
     \hfill
     \begin{subfigure}{0.285\textwidth}
         \includegraphics[width=\linewidth]{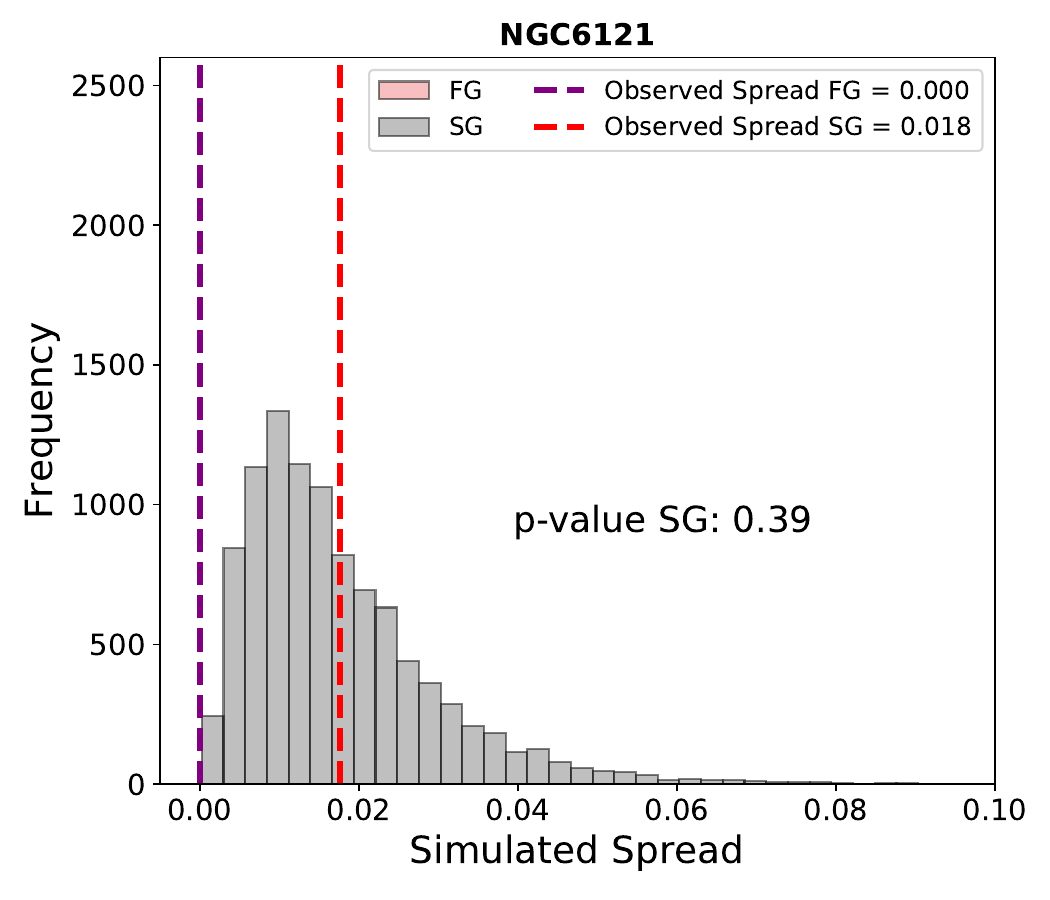}
     \end{subfigure}
     \hfill
     \begin{subfigure}{0.285\textwidth}
         \includegraphics[width=\linewidth]{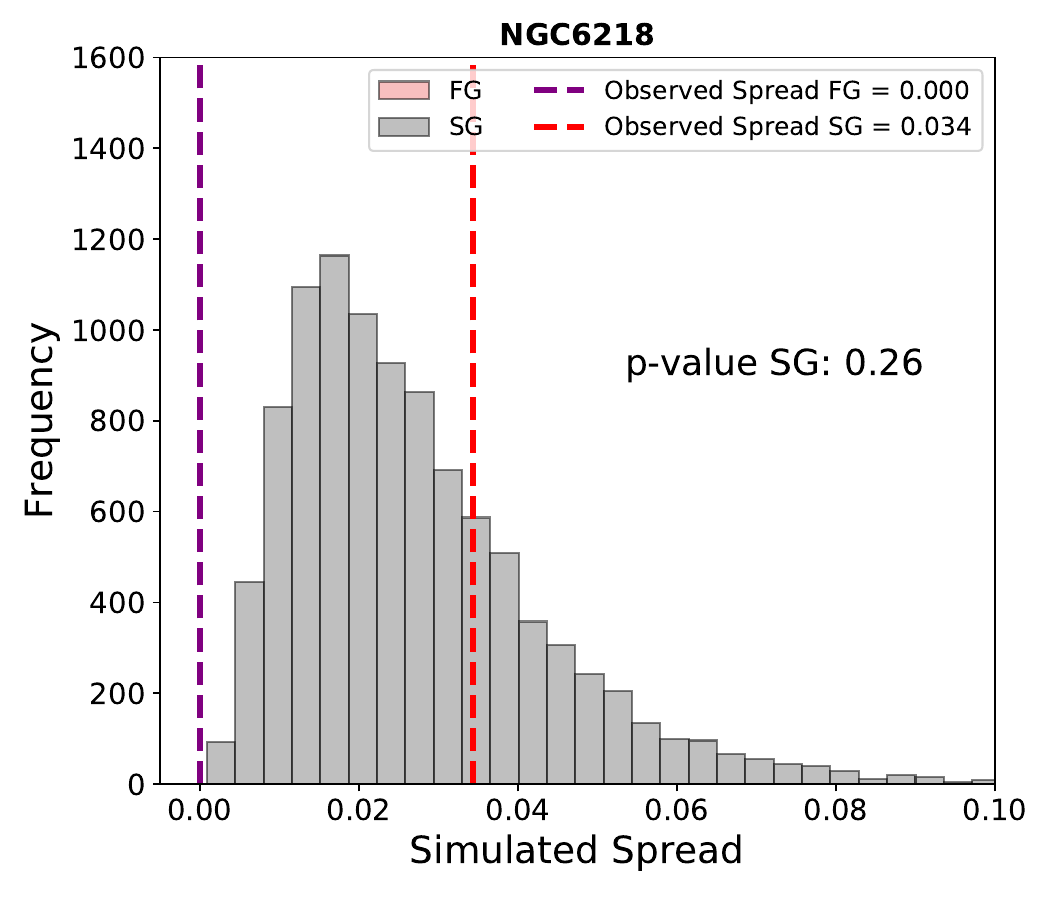}
     \end{subfigure}
     \hfill
     \begin{subfigure}{0.24\textwidth}
         \includegraphics[width=\linewidth]{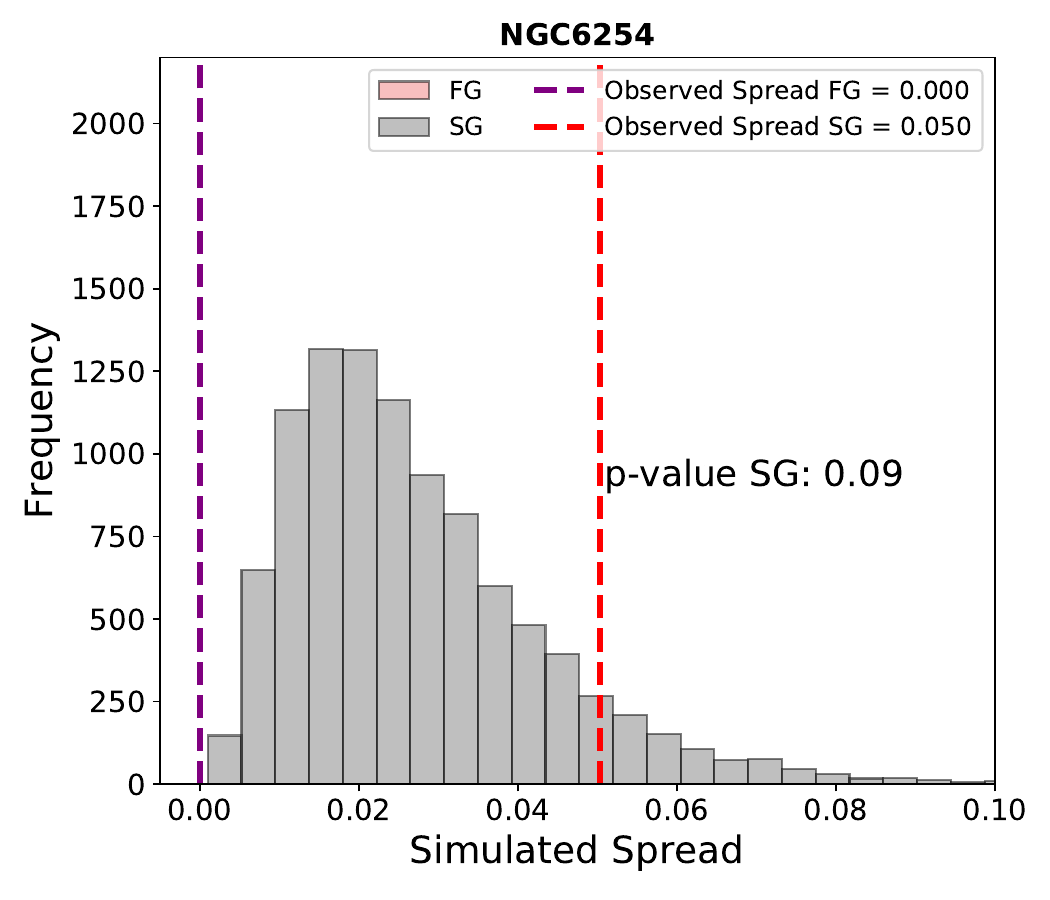}
     \end{subfigure}
     \hfill
     \begin{subfigure}{0.24\textwidth}
         \includegraphics[width=\linewidth]{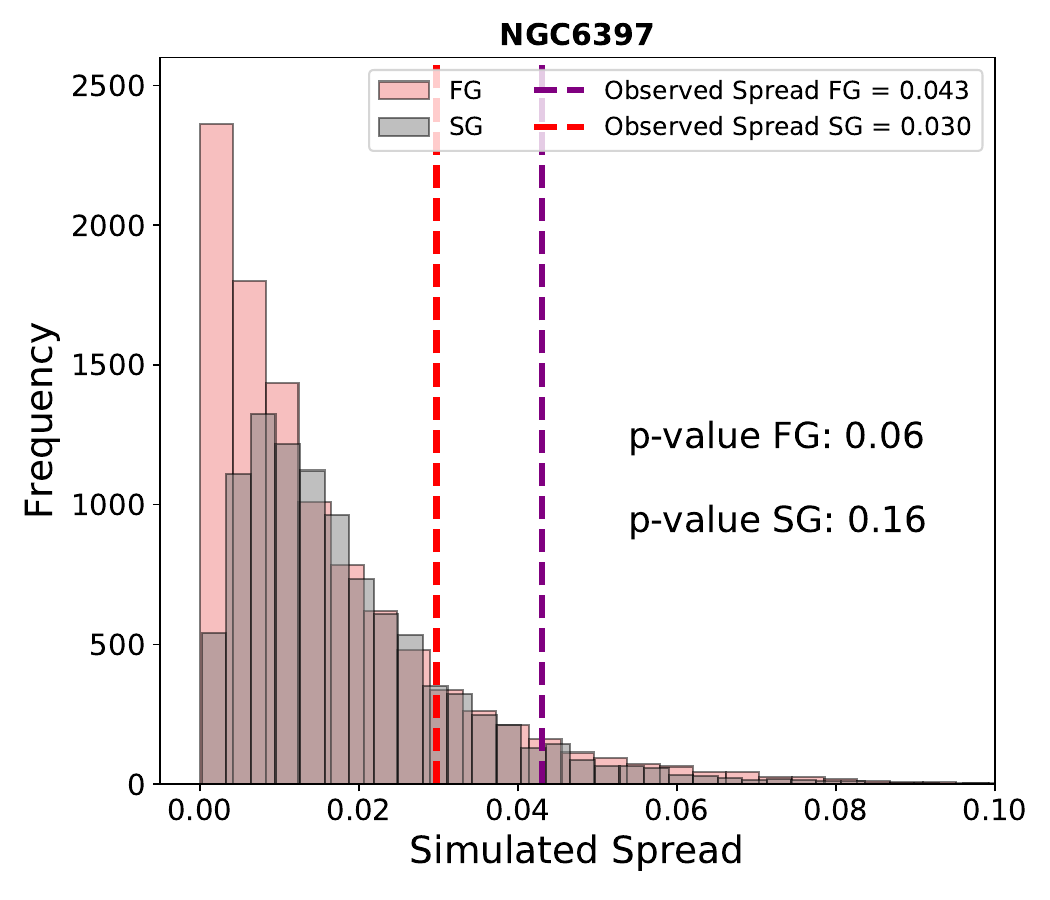}
     \end{subfigure}
     \hfill
     \begin{subfigure}{0.24\textwidth}
         \includegraphics[width=\linewidth]{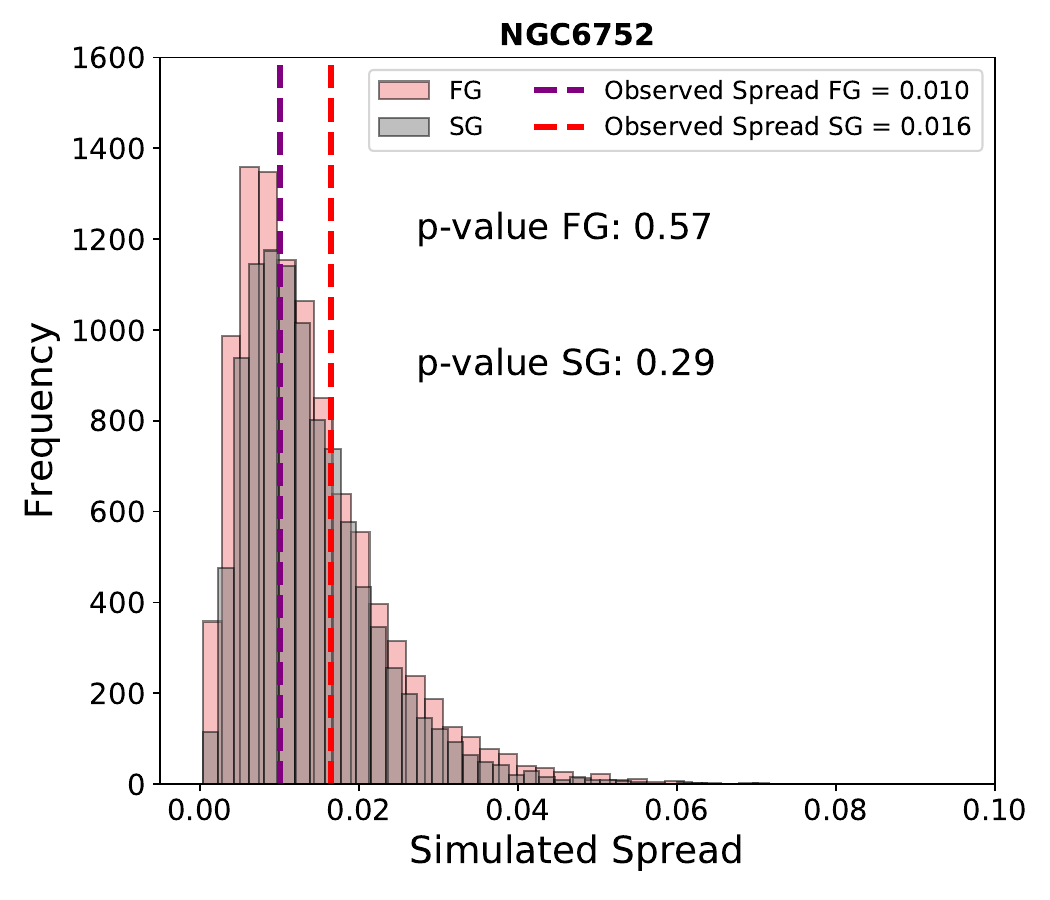}
     \end{subfigure}
     \hfill
     \begin{subfigure}{0.24\textwidth}
         \includegraphics[width=\linewidth]{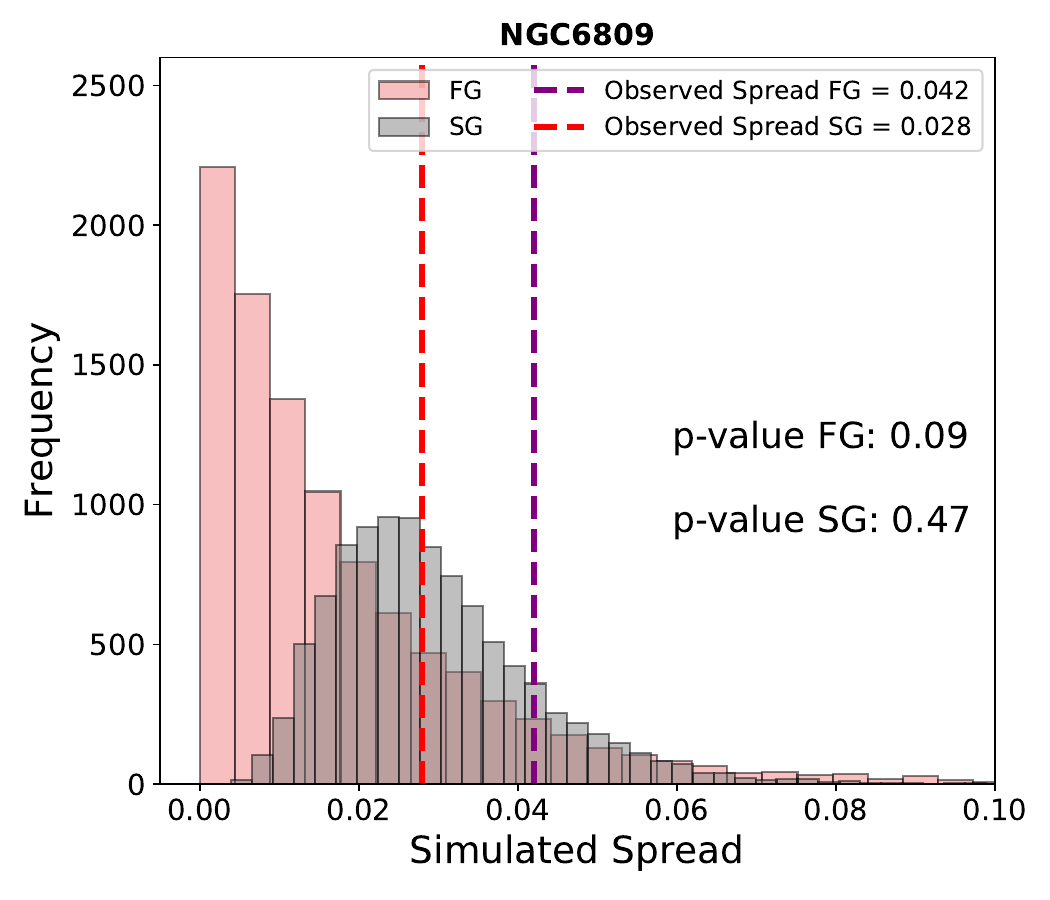}
     \end{subfigure}
     \hfill
     \caption{Histogram of 10,000 Monte simulations of Fe spread by stellar generation. The pink and grey histograms display the distribution of the spread in FG and SG, respectively. Moreover, the dashed lines indicate the observed Fe spread in FG (purple) and SG (red) stars in each GC.}
     \label{fig:FeI_histograms_Gen}
\end{figure*}

Figure~\ref{fig:FeI_histograms_Gen} shows the distributions of simulated spreads obtained from 10,000 Monte Carlo iterations for the two subsamples -- FG and SG stars -- under the assumption of no intrinsic spread beyond the measurement uncertainties. In each panel, pink and grey histograms represent the simulated spreads for FG and SG stars, respectively, while the dashed purple and red lines mark the observed Fe spreads for the corresponding populations.

Three clusters show more pronounced signatures: NGC~1851, NGC~3201, and NGC~5634. Although NGC~1851 does not exhibit an unusually large overall spread compared with other clusters, the spread measured among its SG stars reaches high statistical significance. However, the presence of only one FG star in our sample prevents any meaningful comparison between generations. In NGC~3201, FG stars exhibit an iron spread substantially larger than that measured among SG stars (see Table~\ref{tab:Stats}), and the Monte Carlo simulations indicate a high statistical significance for the FG spread. This result provides strong evidence that the FG population in this cluster displays greater iron variation than the SG population. In NGC~5634, the FG population shows one of the largest iron spreads in the entire sample; however, the presence of only one SG star again prevents any meaningful comparison between generations.

All remaining clusters display iron spreads that remain fully consistent with the measurement uncertainties in both stellar generations. We therefore find no evidence for intrinsic iron variations within either population in those systems.

\begin{table}[!ht]
   \centering
   \caption{Summary table of observed spreads by cluster ($\sigma_{g}$) and stellar generation ($\sigma_{FG}$ and $\sigma_{SG}$).}
   \label{tab:Stats}  
   \resizebox{\columnwidth}{!}{
   \begin{tabular}{cccccccc}
   \hline \hline
        Cluster & $\sigma_{g}$ & $\sigma_{FG}$ & N$_{FG}$ & $\sigma_{SG}$ & N$_{SG}$ & Origin & Type \\ \hline
        NGC~104   & 0.019$\pm{0.010}$ & 0.000 & 1 & 0.011$\pm{0.011}$ & 3  & M-D & I \\ 
        NGC~1851  & 0.038$\pm{0.010}$ & 0.000 & 1 & 0.037$\pm{0.011}$ & 4  & G-E & II\\ 
        NGC~1904  & 0.021$\pm{0.014}$ & 0.000 & 1 & 0.017$\pm{0.014}$ & 6  & G-E & X \\ 
        NGC~2808  & 0.025$\pm{0.010}$ & 0.026$\pm{0.011}$ & 5 & 0.000 & 1  & G-E & I \\ 
        NGC~3201  & 0.069$\pm{0.013}$ & 0.103$\pm{0.021}$ & 3 & 0.016$\pm{0.016}$ & 4  & Seq; G-E & I\\ 
        NGC~5634  & 0.075$\pm{0.017}$ & 0.077$\pm{0.019}$ & 4 & 0.000 & 1  & H99; G-E & X\\ 
        NGC~5904* & 0.035$\pm{0.011}$ & 0.022$\pm{0.014}$ & 2 & 0.038$\pm{0.013}$ & 7  & H99; G-E & I\\ 
        NGC~6121  & 0.017$\pm{0.009}$ & 0.000 & 1 & 0.018$\pm{0.010}$ & 5  & L-E & I\\ 
        NGC~6218* & 0.032$\pm{0.011}$ & 0.000 & 1 & 0.034$\pm{0.011}$ & 6  & M-D & I\\ 
        NGC~6254  & 0.050$\pm{0.013}$ & 0.000 & 1 & 0.050$\pm{0.014}$ & 6  & L-E & I\\ 
        NGC~6397  & 0.037$\pm{0.012}$ & 0.043$\pm{0.021}$ & 2 & 0.030$\pm{0.013}$ & 4  & M-D & I\\ 
        NGC~6752  & 0.024$\pm{0.005}$ & 0.010$\pm{0.008}$ & 4 & 0.016$\pm{0.006}$ & 6  & M-D & I\\ 
        NGC~6809  & 0.041$\pm{0.009}$ & 0.042$\pm{0.023}$ & 2 & 0.028$\pm{0.009}$ & 11 & L-E & I\\ 
        \hline \hline
    \end{tabular}}
    \tablefoot{The columns N$_{FG}$ and N$_{SG}$ report the number of stars assigned to each generation. The origin classification is the one according to \citet{Massari2019}, where M-D, G-E, Seq, H99, and L-E refer to the main disc, Gaia-Enceladus, Sequoia galaxy, Helmi Streams, and an unassociated low-energy GC, respectively. Cluster type classification is based on \citet{Milone2017}. Letter ‘X’ stands for GCs type that were not classified in the mentioned article. The symbol ‘*’ next to NGC~5904 and NGC~6218 indicate that those GCs include two subsamples.}
\end{table}

\section{Discussion}
\label{Sec: Discussion}

The detection of significant internal variations in elements beyond those involved in hot H-burning processes may challenge the current understanding of the multiple stellar population phenomenon in GCs. In our sample, most clusters do not exhibit statistically significant iron spreads, either globally or within individual stellar populations. However, three clusters -- NGC~3201, NGC~1851, and NGC~5634 -- display statistically significant Fe variations. We first discuss these systems individually before comparing our results with previous studies.

\subsection{Clusters with significant iron spreads}

\paragraph{NGC3201}
The case of NGC~3201 has long been debated in the literature. \citet{Simmerer2013} reported a substantial Fe spread of approximately 0.40 dex -- well beyond their observational uncertainties -- suggesting that the cluster may have retained material from SNe Ia despite its relatively low mass. However, \citet{Mucciarelli2015} later attributed the reported spread to non-LTE effects affecting asymptotic giant branch (AGB) stars, implying that the Fe variation is not intrinsic. Our differential approach, which compares sibling stars with similar atmospheric parameters, should minimise non-LTE biases. Therefore, the significant spread detected in our analysis is unlikely to arise from this effect alone. It is also important to note that NGC~3201 exhibits strong differential reddening, which may introduce additional observational uncertainties.
Although NGC~3201 has occasionally been discussed as a candidate cluster hosting multiple metallicity groups, no clear metallicity bimodality is identified in the spectroscopic sample analysed in this work. For this reason, we did not separate the stars into metal-poor and metal-rich subcomponents and instead analysed the cluster as a whole.
Assuming the presence of an intrinsic iron spread in NGC~3201, we performed an additional Monte Carlo analysis similar to that described in the previous section, introducing intrinsic spreads in steps of 0.025 dex. This experiment indicates that the intrinsic dispersion in this cluster is likely between 0.025 and 0.05 dex, consistent with the values reported by \citet{Latour2025}. Evidence in favour of an extragalactic origin may also provide additional context for the peculiar behaviour of NGC~3201. \citet{Massari2019} associated this cluster with the Sequoia or Gaia-Enceladus merger event. However, other Gaia-Enceladus clusters included in our sample (e.g. NGC~1904 and NGC~2808) do not exhibit comparable Fe spreads, suggesting that extragalactic origin alone cannot fully explain the observed behaviour. Nevertheless, since NGC~3201 remains the only cluster in our sample plausibly associated with Sequoia, its origin may still be relevant for understanding its chemical complexity.

\paragraph{NGC~1851}
This GC, which exhibits a modest but statistically significant spread, has attracted considerable attention in previous studies. Through photometric observations, \citet{Milone2008} identified two distinct subgiant branches, suggesting either the merger of two separate GCs or the origin of NGC~1851 within a disrupted dwarf galaxy. The latter scenario agrees with the findings of \citet{Massari2019}, who proposed an extragalactic origin associated with Gaia-Enceladus (see Table~\ref{tab:Stats}). From a spectroscopic perspective, \citet{Carretta2010_n1851} was the first to suggest that the cluster could be the result of a merger of two GCs with slightly different Fe and an age difference of 1 Gyr. Later, similar results were confirmed by \citet{Tautvaivsien2022}. Moreover, \citet{Carretta2011} confirmed an iron spread that exceeded observational uncertainties, indicating possible contributions from supernova enrichment, and later classified this cluster among iron-complex GCs \citep{Johnson2015}. Our detection of a significant iron spread in NGC~1851 therefore agrees with previous studies. However, the presence of only one FG star in our sample prevents an analysis of iron spread within that generation and limits comparisons with the SG population.

\paragraph{NGC~5634}
This GC remains relatively unexplored from a spectroscopic standpoint. Several studies have associated this cluster with an extragalactic origin, although its parent system remains uncertain. Previous works linked NGC~5634 to the Helmi stream \citep{Callingham2022,Massari2019}, Gaia-Enceladus \citep{Massari2019}, or the Sagittarius system \citep{Bellazzini2002}. Among the few available spectroscopic investigations, \citet{Sbordone2015} analysed only two stars, which prevented any statistically meaningful assessment of iron spreads. Later, \citet{Carretta2017} studied seven stars and reported a possible small iron spread based on a comparison between observational uncertainties and the measured dispersion, although they ultimately considered the spread unlikely to be real. As a consistency check, we verified that the stars in common with \citet{Carretta2017} display the same abundance ordering in our analysis, from the lowest to the highest iron abundance. Our results therefore strengthen the evidence for a potentially significant Fe variation in NGC~5634, both at the cluster level and within the FG population.

\subsection{Comparison with previous work}

In a large photometric study of 55 GCs, \citet{Legnardi2022} derived internal iron spreads among FG stars using the colour--metallicity relation of \citet{Dotter2008}, reporting variations of up to 0.30 dex and a mean spread of 0.12 dex. Given that typical spectroscopic detection limits are of the order of 0.10 dex, such results have been questioned in terms of statistical robustness \citep[see e.g.][]{Carretta2025}. We note that the analysis of \citet{Legnardi2022} included main-sequence (MS) stars, whereas our sample consists only of upper RGB stars. However, we do not expect this difference in evolutionary stage to significantly affect the comparison of intrinsic Fe spreads. Any abundance variations produced by atomic diffusion during the MS phase are expected to be largely erased during the first dredge-up.

Our differential approach, based on comparisons between sibling stars, minimises systematics associated with stellar parameters and reduces spurious abundance variations. Within this framework, and with the exception of NGC~3201, none of the clusters in common with \citet{Legnardi2022} shows a highly significant Fe spread among FG stars. It is also important to note that our sample does not include several clusters identified by \citet{Legnardi2022} as exhibiting the largest internal iron spreads (e.g. NGC~2298, NGC~4833, NGC~5024, NGC~5272, NGC~5927, and NGC~7078). These and other systems should be revisited using homogeneous high-resolution spectroscopy, possibly complemented by large surveys such as Gaia-ESO \citep{Pancino2017}, APOGEE \citep{Schiavon2024}, or 4MOST \citep{Lucatello2023}.

We also note that \citet{Dondoglio2025} reported iron spreads comparable to those inferred by \citet{Legnardi2022}. However, their analysis differs from ours in two important respects. First, they compiled abundance measurements from multiple surveys and catalogues rather than deriving abundances through a homogeneous spectroscopic analysis. Secondly, the identification of stellar populations is based on photometric indices primarily sensitive to C and N abundances, whereas our classification relies on Na abundances. These methodological differences may contribute to the contrasting results and complicate a direct comparison between the studies.

Our results are broadly consistent with those of \citet{Latour2025}, who analysed more than 8,000 low-resolution spectra in 21 GCs. Although their sample mainly probes stars along the lower RGB, whereas our analysis focuses on upper RGB stars, we do not expect this difference to significantly affect the inferred intrinsic Fe spreads because effects such as atomic diffusion do not generate significant abundance changes within the rapid RGB evolution.
They reported small Fe dispersions for each cluster, typically below 0.05 dex, and obtained consistent results when separating stellar populations. Notably, in NGC~3201 they did not identify an anomalously large global Fe spread, although  they measured a dispersion approximately 50 percent larger among FG stars than among SG stars.

Finally, \citet{Lardo2023} performed a differential abundance analysis of FG stars in NGC~2808 and reported a significant Fe spread. Although their methodology resembles ours, we did not detect a comparable dispersion in this cluster. The absence of common stars prevents a direct comparison between the two datasets. Notably, their targets are more centrally concentrated, whereas our sample probes more external regions. This spatial difference may contribute to the discrepancy if the iron variations are more pronounced towards the cluster centre. Alternatively, observational effects such as imperfect background subtraction or contamination from neighbouring stars in crowded regions may artificially enhance the measured dispersion. Furthermore, a direct comparison with ChMs would provide additional insights into the discrepancy between our results and those of \citet{Lardo2023}, particularly given the reported sensitivity of the extension of the FG sequence to iron variations. Unfortunately, homogeneous ChM data are not currently publicly available for the full set of clusters analysed in this work. Moreover, existing HST-based photometric samples predominantly cover the central regions of the clusters and do not show spatial overlap with our spectroscopic targets. As a result, we cannot assess whether differences in the ChM distributions of the sampled stars contribute to the contrasting Fe-spread measurements.

\begin{figure}
        \centering
        \includegraphics[width=0.7\columnwidth]{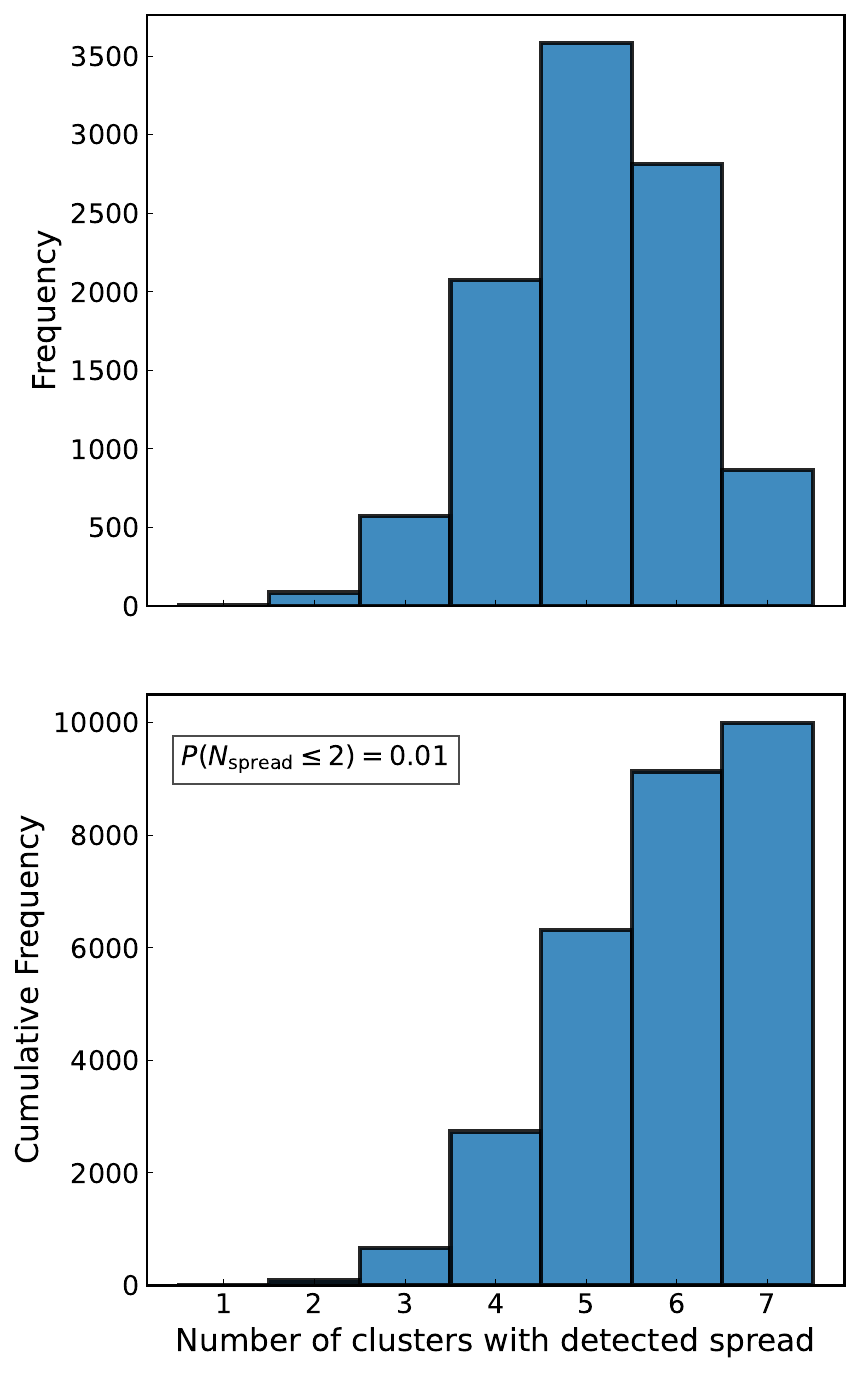}
        \caption{[Upper panel:] Histogram of the number of clusters that showed significant iron spread in their FG after 10,000 Monte Carlo simulations. [Lower panel:] Cumulative distribution of the same simulation, showing a probability of 1\% of getting two or fewer GCs with iron spread in their FG.}
        \label{fig:MC_FG_Legnardi}
\end{figure}

\begin{figure*}
    \sidecaption
        \centering
        \includegraphics[width=12cm]{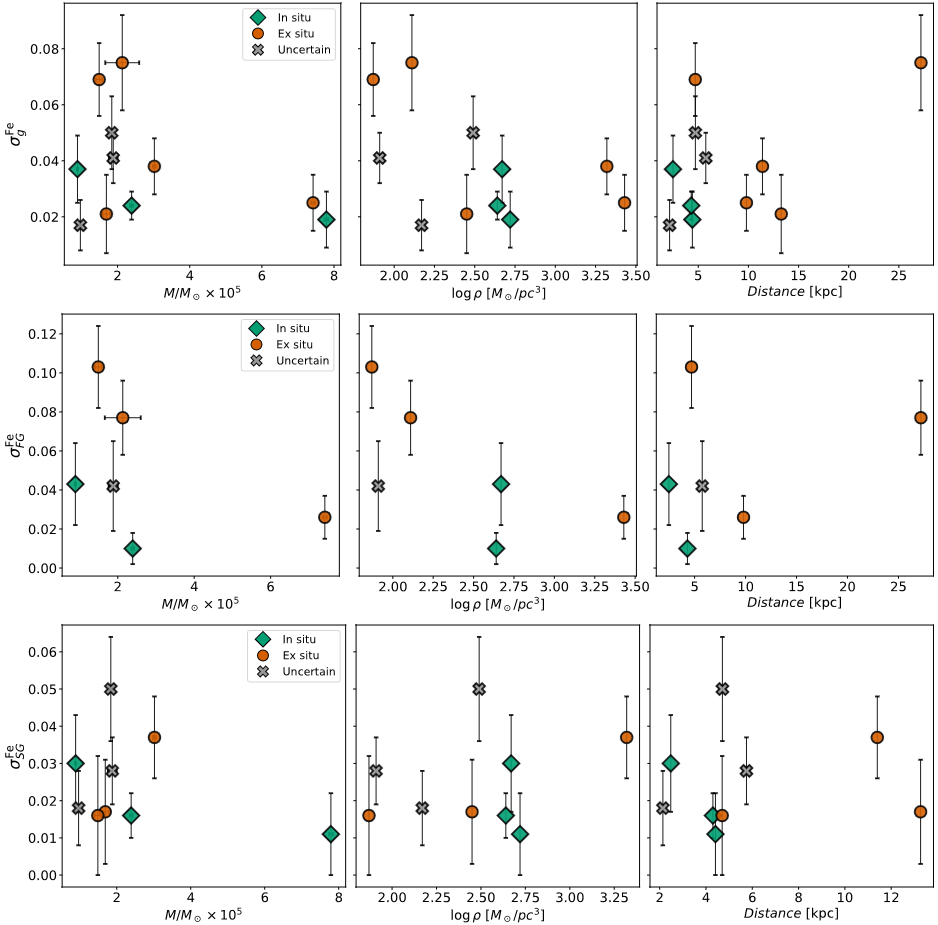}
        \caption{Differential abundance spreads for the full sample (upper panels), FG stars (middle panels), and SG stars (lower panels), shown as a function of cluster mass (left column), stellar density within the half-mass radius (middle column), and distance from the Sun (right column). Symbols are colour-coded by cluster origin: green diamonds for in situ, orange circles for ex situ, and grey crosses for clusters of uncertain origin.}
        \label{fig:globalparameters}
\end{figure*}

\subsection{Monte Carlo validation of literature spreads}

To assess whether the large iron spreads reported by \citet{Legnardi2022} are compatible with our observations, we performed an additional Monte Carlo experiment. We considered the seven clusters in our sample containing at least two FG stars (NGC 2808, NGC 3201, NGC 5634, NGC 5904, NGC 6397, NGC 6752, and NGC 6809), for which an internal FG dispersion can be measured. In each Monte Carlo realisation, we randomly selected one iron dispersion from the distribution of values reported in Table 3 of \citet{Legnardi2022}, spanning 0.00–0.30 dex, and generated a Gaussian abundance distribution centred on zero with the selected dispersion. We then drew a number of stars equal to the observed number of FG stars in each cluster and measured the resulting sample dispersion. We repeated this procedure 10,000 times for each cluster. For each realisation, we recorded whether the simulated dispersion exceeded the observed FG dispersion measured in that cluster. We then counted the number of clusters satisfying this condition in each Monte Carlo iteration. Figure 8 shows the resulting distribution. If the large Fe spreads reported by \citet{Legnardi2022} were representative of the intrinsic FG dispersions of our clusters, we would expect to recover dispersions larger than those observed in most clusters. Instead, obtaining two or fewer clusters with simulated dispersions exceeding the observed values occurs in only 1\% of the Monte Carlo realisations, indicating that the large spreads reported in the literature are difficult to reconcile with our measurements.

\subsection{GC properties and iron spread}

To explore alternative explanations for the observed internal iron variations, we examined several global cluster properties, including origin \citep{Massari2019}, mass, stellar density, and distance from the Sun \citep{Baumgardt2018}. Figure~\ref{fig:globalparameters} compares the measured abundance spreads with these parameters. The figure presents differential abundance spreads for the full sample (upper panels), FG stars (middle panels), and SG stars (lower panels) as functions of cluster mass (left column), stellar density within the half-mass radius (middle column), and distance from the Sun (right column). Symbols are colour-coded according to cluster origin: blue for in situ systems, orange for ex situ systems, and grey for clusters of uncertain origin, following \citet{Massari2019}. Across all panels, we find no clear correlation between abundance spreads and any of the examined global parameters. Spearman correlation tests confirm this conclusion, yielding p values that remain far from statistical significance for all cases.

\section{Summary and conclusions}
\label{Sec: Conclusions}

In this study, we applied a differential abundance analysis to a large sample of Galactic GCs spanning a wide metallicity range. By comparing sibling stars with closely matched stellar parameters, this approach minimises systematic uncertainties associated with classical abundance techniques and provides a high sensitivity to subtle intrinsic iron variations.

Our results show no evidence for significant Fe spreads in the majority of the clusters analysed. This conclusion holds for both the full cluster samples and, where statistics allow, within individual stellar generations. The present data therefore do not support a scenario in which internal iron variations are a common property of Galactic GCs. However, a small number of exceptions emerge. NGC~3201 displays the most significant Fe spread in our sample, with additional tests indicating an intrinsic dispersion of 0.025--0.050 dex. NGC~1851 and NGC~5634 also exhibit statistically significant internal variations, confirming that a limited subset of clusters may have experienced more complex chemical enrichment histories. Our Monte Carlo comparison with the iron spreads among FG stars reported by \citet{Legnardi2022} suggests that such large Fe variations may not be common among GCs. If these spreads were widespread and intrinsic, we would expect a larger fraction of the clusters in our sample to exhibit statistically significant iron variations than we observed.

Overall, these findings indicate that measurable Fe inhomogeneities might be rare and not ubiquitous among Galactic GCs, although larger stellar samples will be required to establish the prevalence of this behaviour across the Galactic GC population. To do so, differential abundance techniques provide a powerful framework for identifying the few systems that preserve signatures of non-standard formation or enrichment pathways.

Future work should extend this methodology to larger stellar samples and additional chemical species. In this context, the next generation of high-resolution multi-object spectrographs \citep{Magrini2023} will be particularly important, enabling precise abundance measurements for large numbers of stars and a more comprehensive reconstruction of the chemical evolution of GCs.

\section*{Data availability}
Table \ref{tab:stellar_parameters} is only available in electronic form at the CDS via anonymous ftp to \url{cdsarc.u-strasbg.fr} (130.79.128.5) or via \url{http://cdsweb.u-strasbg.fr/cgi-bin/qcat?J/A+A/}.

\begin{acknowledgements}
J.S.U. and L.M. acknowledge support from INAF through the Large Grants EPOCH, funding for the WEAVE project, the Mini-Grants Checs (1.05.23.04.02), and financial support under the National Recovery and Resilience Plan (PNRR), Mission 4, Component 2, Investment 1.1, Call for tender No. 104 published on 2 February 2022 by the Italian Ministry of University and Research (MUR), funded by the European Union – NextGenerationEU, through the Project ‘Cosmic POT’ (Grant Assignment Decree No. 2022X4TM3H, MUR).
J.S.U., thanks INAF for its support through the Mini-Grant (1.05.24.07.02). This work was partially funded by the PRIN INAF 2019 grant ObFu 1.05.01.85.14 (\emph{'Building up the halo: chemo-dynamical tagging in the age of large surveys'}, PI. S. Lucatello)      
\end{acknowledgements}

\bibliographystyle{aa}
\bibliography{bibliography}

\end{document}